\documentclass[
 aip,jcp,
 reprint,
 floatfix
]{revtex4-2}

\usepackage{amsmath,amssymb}

\usepackage{changepage}
\usepackage{graphicx}
\usepackage{bm}
\usepackage{hyperref}

\hypersetup{
    colorlinks=true,
    linkcolor=blue,
    citecolor=blue,
    filecolor=blue,
    urlcolor=blue,
}

\usepackage[utf8]{inputenc}
\usepackage[T1]{fontenc}
\usepackage{etoolbox}

\usepackage[looser,scaled=0.951]{newtxtext} 
\usepackage[scaled=0.951]{newtxmath} 

\makeatletter
\def\@email#1#2{%
 \endgroup
 \patchcmd{\titleblock@produce}
  {\frontmatter@RRAPformat}
  {\frontmatter@RRAPformat{\produce@RRAP{*#1\href{mailto:#2}{#2}}}\frontmatter@RRAPformat}
  {}{}
}%
\makeatother

\begin{document}



\title{\Huge Model selection in atomistic simulation}
\author{Jonathan E. Moussa}
\affiliation{Molecular Sciences Software Institute, Virginia Tech, Blacksburg, Virginia 24060, USA}
\email{godotalgorithm@gmail.com}

\begin{abstract}
\begin{adjustwidth}{-1.26cm}{0cm}
There are many atomistic simulation methods with very different costs, accuracies, transferabilities, and numbers of empirical parameters.
I show how statistical model selection can compare these methods fairly, even when they are very different.
These comparisons are also useful for developing new methods that balance cost and accuracy.
As an example, I build a semiempirical model for hydrogen clusters.
\end{adjustwidth}
\end{abstract}

\maketitle

\section{Introduction}

Scientists have been building quantitative atomistic models for over a century \cite{oldest_model}.
In that time, many atomistic models have evolved into sophisticated computer simulations \cite{atomistic_website}.
While there are now models based on a wide variety of atomistic simulation methods, most development has focused on two contradictory goals.
Classical molecular mechanics (MM) methods focus on minimizing cost to access phenomena at large length scales and long time scales \cite{mm_goal_folding}.
However, the use of MM methods is limited by the availability and accuracy of system-specific interatomic potentials \cite{potential_selection}.
In contrast, first-principles quantum mechanics (QM) methods focus on minimizing error for general-purpose simulations \cite{qm_goal}, which can get very expensive.
MM methods can achieve simulation costs of less than $10^{-5}$ CPU-seconds per atom \cite{mm_bench}, while high-accuracy QM methods have asymptotic costs greater than $10^{4}$ CPU-seconds per atom \cite{qm_bench}.

Because of the large gaps in cost and utility, there are many atomistic simulation tasks for which QM methods are too expensive and MM methods have no suitable interatomic potential.
In this situation, a scientist needs an affordable model and must either develop their own or use an existing one such as a semiempirical QM (SQM) model \cite{sqm_review,xtb}.
In either case, they need to collect evidence to support their model.
They must either gather enough reference data to fit a new model, or find enough examples of scientists using an existing model for similar tasks to be confident that it will work for them.
This type of model selection process is a common occurrence in atomistic science, and yet it remains rather informal and subjective much of the time.

In this paper, I advocate for using statistical model selection \cite{model_selection} to develop and compare models for atomistic simulation.
All else being equal, a scientist should fit or choose a model to maximize the probability that they will succeed at their simulation task.
Since the exact probability will be more expensive to compute than the simulation task itself, they must rely on a proxy probability based on related but simpler simulation tasks.
Assumptions about the transferability of a method's accuracy between related simulation tasks are unavoidable in atomistic science.
Also, when considering methods with different numbers of fitting parameters or costs, extra penalties are needed to avoid overfitting or exceeding computational budgets.
These same principles apply to the development of general-purpose models that are intended to be used by many scientists over a broad distribution of simulation tasks.

As an example, I apply statistical model selection to the task of simulating random hydrogen clusters.
First, I generate high-accuracy QM reference data.
Second, I compare the accuracy of some popular SQM models and density functionals from density functional theory (DFT) \cite{dft}.
Third, I build new SQM models by correcting this SQM and QM data with atomic pair potentials.
Here, model selection determines the optimal number of parameters in the pair potentials and the computational budget thresholds for switching between models.

\section{Statistical model selection\label{model_selection}}

The standard practice in fitting atomistic models with parameters is to minimize a distance between model predictions and reference data.
I consider vectors of $m$ reference data points $\mathbf{x}$ and model predictions $\mathbf{y}(\boldsymbol{\lambda})$, which are determined by $n$ real parameters $\boldsymbol{\lambda}$.
The value of $\boldsymbol{\lambda}$ is usually chosen by minimizing the mean absolute error (MAE),
\begin{equation}
\mathrm{MAE}[\mathbf{x} , \mathbf{y}(\boldsymbol{\lambda})]= \frac{\| \mathbf{x} - \mathbf{y}(\boldsymbol{\lambda}) \|_1}{m} = \frac{1}{m} \sum_{i=1}^m | x_i - y_i(\boldsymbol{\lambda})|,
\end{equation}
 or the root-mean-square deviation (RMSD),
\begin{equation} \label{rmsd}
 \mathrm{RMSD}[\mathbf{x} , \mathbf{y}(\boldsymbol{\lambda})] = \frac{\| \mathbf{x} - \mathbf{y}(\boldsymbol{\lambda}) \|_2}{\sqrt{m}} = \sqrt{ \frac{1}{m} \sum_{i=1}^m [ x_i - y_i(\boldsymbol{\lambda})]^2}.
\end{equation}
The general expectation is that smaller distances correspond to better accuracy and thus a higher chance of success when these models are used for other simulation tasks.
However, this relationship is indirect because these distances are not operational measures of success.
An operational measure would describe the application of a model by scientists in a more explicit and direct way, including how successful they are.
Directly optimizing an operational measure should produce a more successful model if the operational measure itself is sufficiently accurate.
To use statistical model selection as an operational measure in this context, I must first introduce two distinct sources of randomness.

The first source of randomness is in the model predictions.
I consider a generalization of the reference information from data points $\mathbf{x}$ to simulation tasks $\mathbf{X}$.
Each reference simulation task $X_i$ defines one or more physical systems and calculations to perform, together with reference output data and success criteria.
The conditional probability of success, $p(\boldsymbol{\lambda} | X_i)$, after choosing a task $X_i$ and using a model with parameters $\boldsymbol{\lambda}$ replaces a distance between $x_i$ and $y_i(\boldsymbol{\lambda})$.
The only constraint on the success criteria is that the success probability for the method used to generate the reference data must be one.
Viable models must always have a nonzero success probability, which requires the model output or success criteria to have a random component.

The second source of randomness is in the choice of reference simulation tasks.
I relate a set of reference simulation tasks to the actual simulation task that a scientist wants to succeed at by considering them to be randomly drawn from a common distribution of simulation tasks.
The probability of choosing a simulation task $X$ is $p(X)$, and the probability of choosing this task and then succeeding with the model is
\begin{equation}
 p(\boldsymbol{\lambda}, X) = p(\boldsymbol{\lambda} | X) p(X).
\end{equation}
It is not strictly necessary for the simulation tasks to have been randomly drawn from this distribution.
Such a distribution is still formally useful even when it is an artificial context and not even precisely defined.
It is simply the mathematical representation of a computational scientist as a distribution over simulation tasks.

\subsection{Maximum likelihood estimation}

I now apply the framework of maximum likelihood estimation (MLE) \cite{model_selection} to determine the best model in this randomized setting.
The operational measure of modeling success is the probability of succeeding at all $m$ reference simulation tasks,
\begin{equation} \label{success_probability}
 P(\boldsymbol{\lambda}) = \prod_{i=1}^m p(\boldsymbol{\lambda} | X_i).
\end{equation}
This formula assumes that the random component determining the success or failure of each simulation task is statistically independent from other tasks.
It is related to a statistical likelihood function,
\begin{equation}
 L(\boldsymbol{\lambda}) = \prod_{i=1}^m p(\boldsymbol{\lambda} , X_i) = P(\boldsymbol{\lambda}) \prod_{i=1}^m p(X_i),
\end{equation}
 over the joint distribution of simulation tasks and modeling success or failure events.
I follow the common convention of considering the negative logarithm of the probability or likelihood,
\begin{align}
 \hat{D}(\boldsymbol{\lambda}) = - \log P(\boldsymbol{\lambda}) &= - \sum_{i=1}^m \log p(\boldsymbol{\lambda} | X_i) \notag \\
  &= - \log L(\boldsymbol{\lambda}) + \sum_{i=1}^m \log p(X_i),
\label{success_measure}
\end{align}
 which replaces the product over reference simulation tasks with a more convenient sum.
The negative logarithm is a strictly monotonically decreasing function, and maximizing it corresponds to maximizing the likelihood. 
Since $p(X_i)$ has no dependence on $\boldsymbol{\lambda}$, $P(\boldsymbol{\lambda})$ and $L(\boldsymbol{\lambda})$ are maximized by the same value of $\boldsymbol{\lambda}$.

Eq.\@ (\ref{success_measure}) can be interpreted as a statistical estimator of
\begin{equation}
 D(\boldsymbol{\lambda}) = - m \sum_{X} p(X) \log p(\boldsymbol{\lambda} | X),
\label{divergence}
\end{equation}
 which is $m$ times the Kullback-Leibler divergence \cite{kl_divergence} of the always successful reference distribution from the model distribution that can fail at simulation tasks.
The asymptotic success probability for $r \gg 1$ tasks drawn from $p(X)$ is $\exp[- r D(\boldsymbol{\lambda})/m]$ \cite{kl_divergence}, 
 just as $\hat{D}(\boldsymbol{\lambda})$ corresponds to a specific success probability, $\exp[-\hat{D}(\boldsymbol{\lambda})]$.

The individual success probabilities $p(\boldsymbol{\lambda} | X_i)$ are a flexible concept with many possible choices.
I provide a specific example here that is used for the hydrogen cluster example in Sec.\@ \ref{hydrogen_example}, but other choices are discussed in Sec.\@ \ref{probability_choices}.
I also show how the example reduces to the familiar case of RMSD from Eq.\@ (\ref{rmsd}) as a limiting approximation.

A simple example of a success criterion for a task $X_i$ is that a model output $y_i(\boldsymbol{\lambda})$ is within some distance $\epsilon$ of a reference value $x_i$ \cite{success_probability}.
To guarantee a finite success probability and statistical independence, I add a Gaussian random variable $z$ to $y_i(\boldsymbol{\lambda})$ with mean $\mu$ and standard deviation $\sigma$.
The success criterion is then
\begin{equation} \label{success_target}
  x_i - \epsilon \le y_i(\boldsymbol{\lambda}) + z \le x_i + \epsilon,
\end{equation}
 and the probability of it being satisfied is
\begin{align}
 p(\boldsymbol{\lambda} | X_i) &= \int_{x_i - y_i(\boldsymbol{\lambda}) - \epsilon}^{x_i - y_i(\boldsymbol{\lambda}) + \epsilon} \frac{e^{-(z - \mu)^2/(2\sigma^2)}}{\sigma \sqrt{2 \pi}} d z \notag \\
&= \tfrac{1}{2} \mathrm{erf}\left(\tfrac{x_i - y_i(\boldsymbol{\lambda}) - \mu + \epsilon}{\sqrt{2} \sigma}\right) - \tfrac{1}{2} \mathrm{erf}\left(\tfrac{x_i - y_i(\boldsymbol{\lambda}) - \mu - \epsilon}{\sqrt{2} \sigma}\right) .
\label{exact_success}
\end{align}
While not explicit in this notation, error model parameters such as $\mu$ and $\sigma$ are included in the overall parameter vector $\boldsymbol{\lambda}$.
I also consider quantities with strict inequalities on predicted values such as electron addition and removal energies.
Within a vacuum, the energy to add an electron cannot be greater than zero, and the energy to remove an electron cannot be less than zero.
A modified success criterion for a non-negative prediction would be
\begin{equation}
  x_i - \epsilon \le \max\{ y_i(\boldsymbol{\lambda}) + z , 0 \} \le x_i + \epsilon,
\end{equation}
 and its modified probability of being satisfied is
\begin{equation}
 p(\boldsymbol{\lambda} | X_i) = \tfrac{1}{2} + \tfrac{1}{2} \mathrm{erf}\left(\tfrac{x_i - y_i(\boldsymbol{\lambda}) - \mu + \epsilon}{\sqrt{2} \sigma}\right)
\label{signed_success}
\end{equation}
 for $x_i \le \epsilon$ and Eq.\@ (\ref{exact_success}) otherwise.
Non-positive predictions can be modified in a similar way.

The familiar case of RMSD follows from a simplification of Eq.\@ (\ref{exact_success}) for $\sigma \gg \epsilon$.
In this high-variance regime, Eq.\@ (\ref{exact_success}) can be interpreted as a finite-difference approximation to the derivative of the error function,
\begin{equation}
 - \log p(\boldsymbol{\lambda} | X_i) \approx \frac{[x_i - y_i(\boldsymbol{\lambda}) - \mu]^2}{2\sigma^2} + \frac{1}{2} \log \frac{\pi \sigma^2}{2\epsilon^2} + O\left(\frac{\epsilon}{\sigma}\right).
\label{success_probability_approx}
\end{equation}
This is not always a good approximation because it can be negative whereas any exact $- \log p(\boldsymbol{\lambda} | X_i)$ must be strictly non-negative.
If all tasks are assigned $\mu = 0$ and a common $\sigma$ value, then the approximate success measure is a function of the RMSD,
\begin{equation} \label{single_gaussian_mle}
  \hat{D}(\boldsymbol{\lambda}) \approx  \frac{m \, \mathrm{RMSD}[\mathbf{x} , \mathbf{y}(\boldsymbol{\lambda})]^2}{2 \sigma^2} + \frac{m}{2} \log \frac{\pi \sigma^2}{2\epsilon^2}.
\end{equation}
The minimizing $\sigma$ value is simply the RMSD itself, and the minimum value of the approximate success measure is
\begin{equation} \label{rmsd_connection}
 \hat{D}(\boldsymbol{\lambda}) \approx \frac{m}{2} + \frac{m}{2} \log \frac{\pi \, \mathrm{RMSD}[\mathbf{x} , \mathbf{y}(\boldsymbol{\lambda})]^2}{2\epsilon^2}.
\end{equation}
In this scenario, the RMSD and success measure both produce the same minimizing models and rank them in the same order.

\subsection{Information criteria\label{param_penalty}}

Simple MLE is capable of selecting the best model from one family of models parameterized by $\boldsymbol{\lambda}$, but it cannot reliably compare models from different families.
Adding more free parameters to an existing model and optimizing them can only improve the success measure, and nested models with more parameters will always be preferred.
This can eventually cause the modeling phenomenon of overfitting whereby finite-sampling artifacts in the reference data distort the parameter optimization process.
Additional modeling criteria are needed to eliminate parameters that are not useful.
The most common approach is to introduce a penalty for adding model parameters that is overcome by useful parameters.
Objective functions with parameter penalties are usually referred to as information criteria in statistical model selection.
Here, I consider information criteria that are penalties to $\hat{D}(\boldsymbol{\lambda})$ in Eq.\@ (\ref{success_measure}) of the form $\hat{D}(\boldsymbol{\lambda}) + \Delta$.
This deviates slightly from convention in the statistics literature, which usually considers objective functions of the form $-2 \log L (\boldsymbol{\lambda}) + 2 \Delta$.

The simplest, oldest, and most famous penalty is from the Akaike information criterion (AIC) \cite{aic},
\begin{equation}
 \Delta_{\mathrm{AIC}} = n,
\label{aic_bias}
\end{equation}
 where $n$ is the number of parameters in the model.
An important assumption made by the AIC is that the optimized model has success probabilities close to one.
This assumption is relaxed by the penalty of the Takeuchi information criterion (TIC) \cite{tic},
\begin{align}
 \Delta_{\mathrm{TIC}} &= \mathrm{tr} [ \mathbf{F}^{-1} \mathbf{\tilde{F}} ], \ \ \ \ \ \ [\mathbf{F}]_{i,j} =  \frac{\partial^2 \hat{D}}{\partial \lambda_i \partial \lambda_j} (\boldsymbol{\lambda}^*), \notag \\
  [\mathbf{\tilde{F}}]_{i,j} &= \sum_{k=1}^m \left[ \frac{\partial \log p(\boldsymbol{\lambda} | X_k)}{\partial\lambda_i} \frac{\partial \log p(\boldsymbol{\lambda} | X_k)}{\partial\lambda_j} \right]_{\boldsymbol{\lambda} = \boldsymbol{\lambda}^*}, 
\label{tic_bias}
\end{align}
 where $\boldsymbol{\lambda}^*$ is the minimizer of $\hat{D}(\boldsymbol{\lambda})$.
The AIC penalty is a good approximation of the TIC penalty when $\mathbf{\tilde{F}} \approx \mathbf{F}$.
In this regime, $\mathbf{\tilde{F}}$ and $\mathbf{F}$ are approximately $m$ times the Fisher information matrix \cite{kl_divergence} of $p(\boldsymbol{\lambda} , X)$.

I provide a derivation of the AIC and TIC penalties that is accessible to physical scientists in Appendix \ref{ic_derivation}.
The derivation makes two assumptions.
The first assumption is that $\hat{D}(\boldsymbol{\lambda})$ and $D(\boldsymbol{\lambda})$ are twice differentiable with respect to $\boldsymbol{\lambda}$ near $\boldsymbol{\lambda}^*$.
The second assumption is that there is enough reference data for the second derivatives of $\hat{D}(\boldsymbol{\lambda})$ and $D(\boldsymbol{\lambda})$ to be good approximations of each other at $\boldsymbol{\lambda}^*$.

\subsection{Cost penalties\label{cost_penalties}}

The primary purpose of fitting models in statistics is to explain data in the absence of a prior explanation.
In contrast, the purpose of fitting models for atomistic simulation is to avoid the large cost of evaluating a known first-principles model.
Statistics is concerned with efficiency, but its main consideration is in getting the most value out of limited data to avoid the potentially high cost of collecting or generating data.
Without some penalty for the cost of models, the inevitable conclusion of statistical model selection in atomistic simulation is to choose the expensive model that was used to generate the reference data.
However, it is useful to have a penalty that disappears when the target computational budget is achieved to preserve the structure of statistical model selection for models that match the budget.

I treat this as a multi-objective optimization problem \cite{moo} with two objectives: computational cost and modeling success.
Using the linear scalarization approach to multi-objective optimization, I combined the objectives as
\begin{equation} \label{multi_objective}
 \hat{D}(\boldsymbol{\lambda}) + \gamma (t - t_0),
\end{equation}
 where $\gamma$ is a Lagrange multiplier, $t$ is the total cost of applying the model to the $m$ simulation tasks, and $t_0$ is the target computational budget.
For $\gamma > 0$, this combined objective penalizes models with $t > t_0$ and rewards models with $t < t_0$.
The cost constraint is formally satisfied by requiring the combined objective to be stationary with respect to $\gamma$,
\begin{equation}
  \frac{\partial}{\partial \gamma} \left[ \hat{D}(\boldsymbol{\lambda}) + \gamma (t - t_0) \right] = t - t_0 = 0.
\end{equation}
However, it is not typical for specific model families to have a highly adjustable cost, and this stationary condition often cannot be satisfied in practice.

The statistical nature of model selection provides an alternate strategy for satisfying cost constraints by mixing models.
I consider a scenario with two models having success measure values $\hat{D}_1$ and $\hat{D}_2$ for costs $t_1$ and $t_2$ such that $\hat{D}_1 > \hat{D}_2$ and $t_1 < t_0 < t_2$.
I then define a hybrid model that randomly selects one of these two models with probabilities
\begin{equation} \label{mix_p}
 p_1 = \frac{t_2 - t_0}{t_2 - t_1} \ \ \ \mathrm{and} \ \ \ p_2 = \frac{t_0 - t_1}{t_2 - t_1}
\end{equation}
 for each task, which achieves an expected cost of $t_0$.
The expected success measure of the hybrid model is more complicated, but it can be bounded from above with Jensen's inequality as
\begin{equation} \label{mix_success}
 p_1 \hat{D}_1 + p_2 \hat{D}_2.
\end{equation}
Any parameter penalties from Sec.\@ \ref{param_penalty} for the two models, $\Delta_1$ and $\Delta_2$, combine as $\Delta_1 + \Delta_2$ for the hybrid model, which acts as a penalty for mixing models.
Thus any computational budget can be matched by combining a cheap, inaccurate model with an expensive, accurate model.

\subsection{Model optimization}

The complete scoring function for a model family including both parameter penalties from Sec.\@ \ref{param_penalty} and cost penalties from Sec.\@ \ref{cost_penalties} is
\begin{equation} \label{full_objective}
 \hat{D}(\boldsymbol{\lambda}^*) + \Delta + \gamma (t - t_0),
\end{equation}
 where $\boldsymbol{\lambda}^*$ is a local minimizer of $\hat{D}(\boldsymbol{\lambda})$ and ideally its global minimizer.
This prescription assumes that costs do not depend sensitively on the choice of $\boldsymbol{\lambda}$.
Otherwise, the minimization of $\hat{D}(\boldsymbol{\lambda})$ and the derivation of parameter penalties in Appendix \ref{ic_derivation} need to include cost penalties and their $\boldsymbol{\lambda}$ derivatives.
The need for $\boldsymbol{\lambda}$ derivatives might also require the use of analytical cost models rather than observed costs.

An important implicit assumption of model selection is a practical ability to minimize $\hat{D}(\boldsymbol{\lambda})$ over $\boldsymbol{\lambda}$.
This includes adjusting values of $\boldsymbol{\lambda}$ to find local minimizers and choosing initial values for $\boldsymbol{\lambda}$ in the basin of convergence for the global minimizer.
Ideally, both first and second $\boldsymbol{\lambda}$ derivatives of $\hat{D}(\boldsymbol{\lambda})$ are available during minimization to use Newton's method and achieve quadratic local convergence.
If only first derivatives are available, local optimization may be restricted to quasi-Netwon methods \cite{quasinewton}, and Eq.\@ (\ref{tic_bias}) may need finite-difference approximations of second derivatives.
While there is not enough generic structure to guarantee or verify global minima, there are often physical considerations to guide reasonable choices of initial $\boldsymbol{\lambda}$ values.
Local minima are still valid model choices in Eq.\@ (\ref{full_objective}), and the selection of local minima is comparable to the selection of model families.

The outer layer of model selection is the decision of which model families to optimize and compare.
In the simplest cases, there are either a small number of model families that can be exhaustively compared or an ordered sequence that can be compared until a clear minimum of Eq.\@ (\ref{full_objective}) is observed.
In general, a model family can be defined by multiple, independent design decisions that cause a combinatorial growth of possible model families.
To avoid the combinatorial growth of model selection costs, a greedy selection process could be used that scores each design decision independently.
These decisions may not actually be independent, and a combinatorial selection process can be systematically approached by considering clusters of simultaneous design decisions and increasing the cluster size.

\section{Hydrogen cluster example\label{hydrogen_example}}

To demonstrate the principles of statistical model selection, I consider a simple set of simulation tasks on randomly generated hydrogen clusters.
By only considering hydrogen atoms, I keep the elemental diversity at a minimum to simplify the process of fitting SQM models with element-specific parameters.
I keep the phenomenological diversity high by considering two distributions of clusters.
A ``dense'' distribution of clusters forces the minimum interatomic distance between hydrogen atoms to be less than the Coulson-Fischer point \cite{coulson_fischer} near 1 {\AA}, while a ``sparse'' distribution allows larger minimum separations.
Molecular orbitals tend to remain grouped into pairs with opposite spin and similar spatial character in the dense distribution, while the sparse distribution generates many clusters that favor spin-polarized, atom-localized orbitals.
Because of limitations in methods and software that generate accurate and reliable reference data, the only observable that I consider is the total energy of clusters.
I consider three simulation tasks to calculate energies for three cluster modifications: removal of an atom, removal of an electron, and addition of an electron.
A success is defined as the calculation of one such energy with an error of 1 kcal/mol or less.
While these distributions and tasks are artificial and not directly motivated by any application, there is some experimental interest in positively \cite{cation_cluster} and negatively \cite{anion_cluster} charged hydrogen clusters.

I generate the dense and sparse distribution of hydrogen clusters by sequential rejection sampling.
Atoms are assigned uniformly random positions in a box containing the valid domain, and the atom is rejected and repositioned if it violates a distance constraint.
The minimum allowed interatomic distance for both distributions is 0.3 {\AA}, near the classical turning point of the H$_2$ potential energy surface.
The maximum allowed value for the minimum interatomic distance is 1 {\AA} for the dense distribution and 4 {\AA} for the sparse distribution.
The sparse distribution is a strict superset of the dense distribution, and some sparse clusters could be recycled as dense clusters.
However, the recycling rate of two-atom clusters is only $0.016$, and it decreases rapidly with increasing cluster size.
This is an example of low recycling efficiency between distributions that are very different.
For the reference data set, I generate 10,000 nested sequences of clusters between two and seven atoms for each distribution, resulting in 120,000 distinct structures.
This nesting structure allows atom removal energies to be extracted from total energies.
The three simulation tasks require calculations of three different charge states -- 0, 1, and -1 -- corresponding to 360,003 total energy calculations including an isolated hydrogen atom.

\begin{figure}
\includegraphics{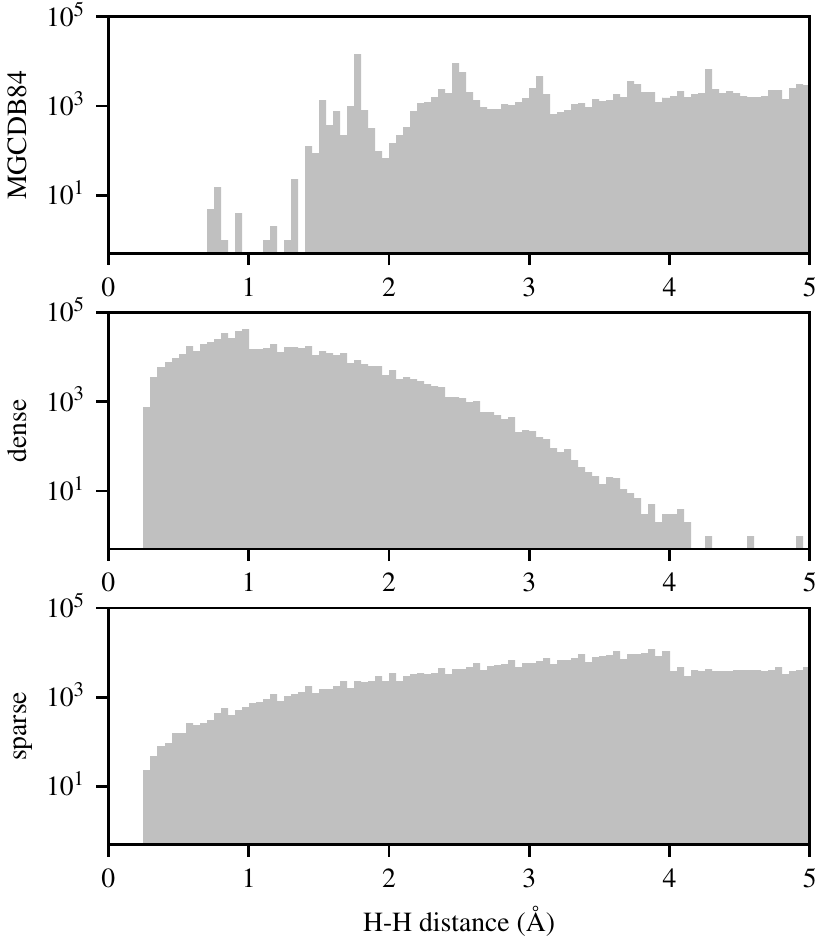}
\caption{Histograms of interatomic distances between hydrogen atoms in the structures from three reference data sets.}
\label{h2_distances}
\end{figure}

\begin{figure*}[!h]
    \centering
    \includegraphics{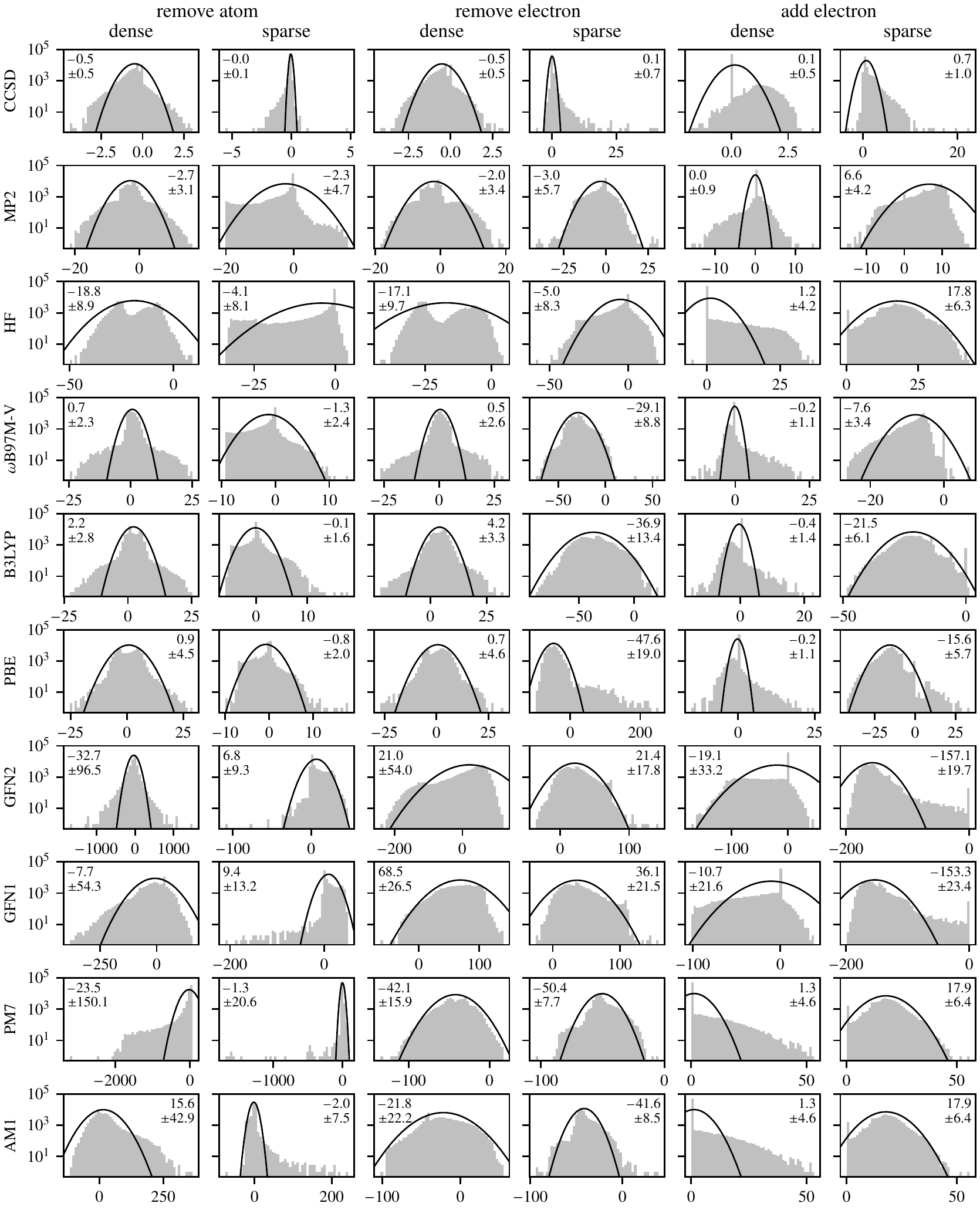}
    \caption{Error histograms in kcal/mol for all models and tasks along with their means, standard deviations, and moment-matching Gaussian model fits.}
    \label{error_plot}
\end{figure*}

\begin{figure*}[!t]
    \centering
    \includegraphics{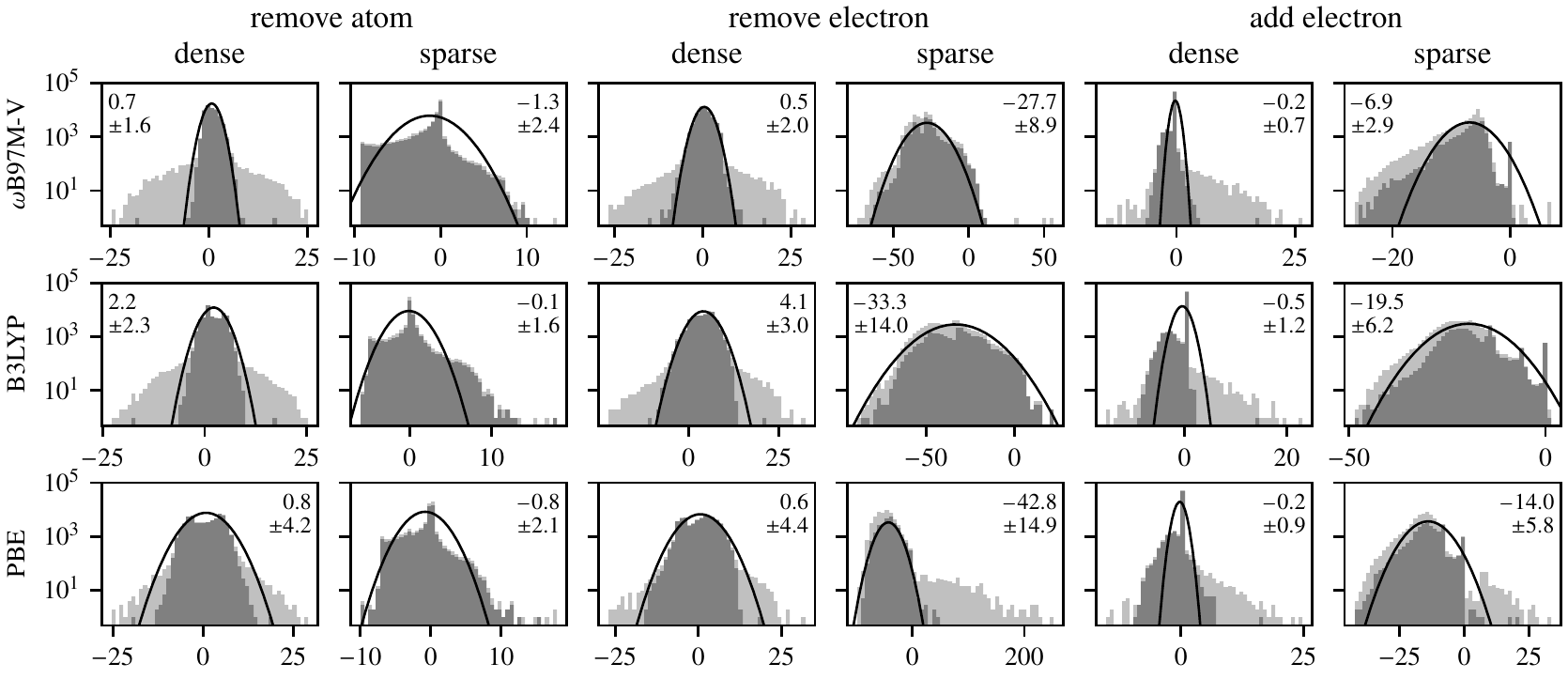}
    \caption{Error histograms in kcal/mol for DFT models and all tasks along with the means, standard deviations, and moment-matching Gaussian model fits of the marked data with consistent total spin values between HF and DFT.}
    \label{spin_marked_plot}
\end{figure*}

I briefly compare this reference data set with the MGCDB84 data set that is popular for testing DFT functionals \cite{mgcdb84}.
Both data sets are restricted to total energies of small, isolated groups of atoms.
MGCDB84 corresponds to 5,931 total energy calculations of structures that are 52.6\% hydrogen, 29.2\% carbon, 8.8\% oxygen, 5.5\% nitrogen, and less than 1\% each of main-group elements from the first four rows of the periodic table.
Thus, while it is not restricted to only hydrogen atoms, hydrogen is the most well-represented element in MGCDB84.
MGCDB84 is organized into 84 subsets of data corresponding to different simulation tasks, including non-covalent binding energies, isomerization energies, formation energies, and barrier heights.
However, this data set lacks diversity by some measures, such as 95.2\% of the structures being closed-shell singlets and 93.0\% being charge neutral.
Also, MGCDB84 mostly contains structures and properties of interest to organic chemistry with structures at equilibrium or saddle points.
MGCDB84 is thus a reasonable proxy for the interests of organic chemists, while the hydrogen cluster data sets broadly sample from the potential energy surface of many hydrogen atoms.
Of particular interest when fitting distance-dependent parameters such as pair potentials and one-body matrix elements is the distribution of interatomic distances between hydrogen atoms.
These distance distributions are shown in Fig.\@ \ref{h2_distances} for MGCDB84 and the two distributions of hydrogen clusters considered here.
MGCDB84 has poor coverage at distances less than 1.4 {\AA} and is not a good reference to fit distance-dependent parameters for hydrogen interactions.

\subsection{Reference data}

I gathered high-level reference data for the hydrogen clusters at the CCSD(T) level of theory \cite{ccsdt} with the def2-QZVPP basis set \cite{qzvpp}.
I also recorded data at the Hartree-Fock (HF), MP2, and CCSD levels of theory during the CCSD(T) calculations.
In addition to the high-level reference data, I also gathered data using several popular SQM models and DFT functionals to test their transferability.
There are too many SQM models and DFT functionals to test all of them, and this study was limited to a few important representative examples.
AM1 \cite{am1} was the most popular SQM thermochemistry model of the last century, and PM7 \cite{pm7} is the most recent model from that family of MNDO-like models \cite{mndo}.
GFN1 \cite{gfn1} and GFN2 \cite{gfn2} are two recent SQM models from the density functional tight-binding (DFTB) framework \cite{dftb}.
PBE \cite{pbe} is the most popular DFT functional in solid-state physics and materials science.
B3LYP \cite{b3lyp} is the most popular DFT functional in chemistry.
$\omega$B97M-V \cite{wb97mv} is claimed to be the most accurate DFT functional without including terms from many-body perturbation theory.
While a smaller basis set might be sufficient, I performed all DFT calculations using the def2-QZVPP basis set for consistency.
In total, I gathered data from eleven QM and SQM models, corresponding to 3,960,033 total energy calculations.

Details of the reference data generation process are presented in Appendix \ref{data_generation}.
This was a challenging, large-scale computational task that required a substantial amount of automation.
Not every calculation was successful, and it was not possible to fix failed calculations by hand at this scale.
The main source of heralded failures was convergence failures of the direct inversion in the iterative subspace (DIIS) algorithm \cite{diis} and related iterative solvers.
The likely main source of silent failures was false convergence of DIIS-based algorithms to metastable states instead of ground states.
While strong electron correlation effects can occur in hydrogen clusters \cite{hydrogen_chain}, I performed multi-reference tests on every structure and did not observe any strong correlations.
Many hydrogen clusters have a frustrated energy landscape, and a robust ground state search required the self-consistent field (SCF) cycle to be initialized with a full set of spin-polarized atomic guesses.

\subsection{Anomaly detection}

\begin{figure*}[!t]
    \centering
    \includegraphics{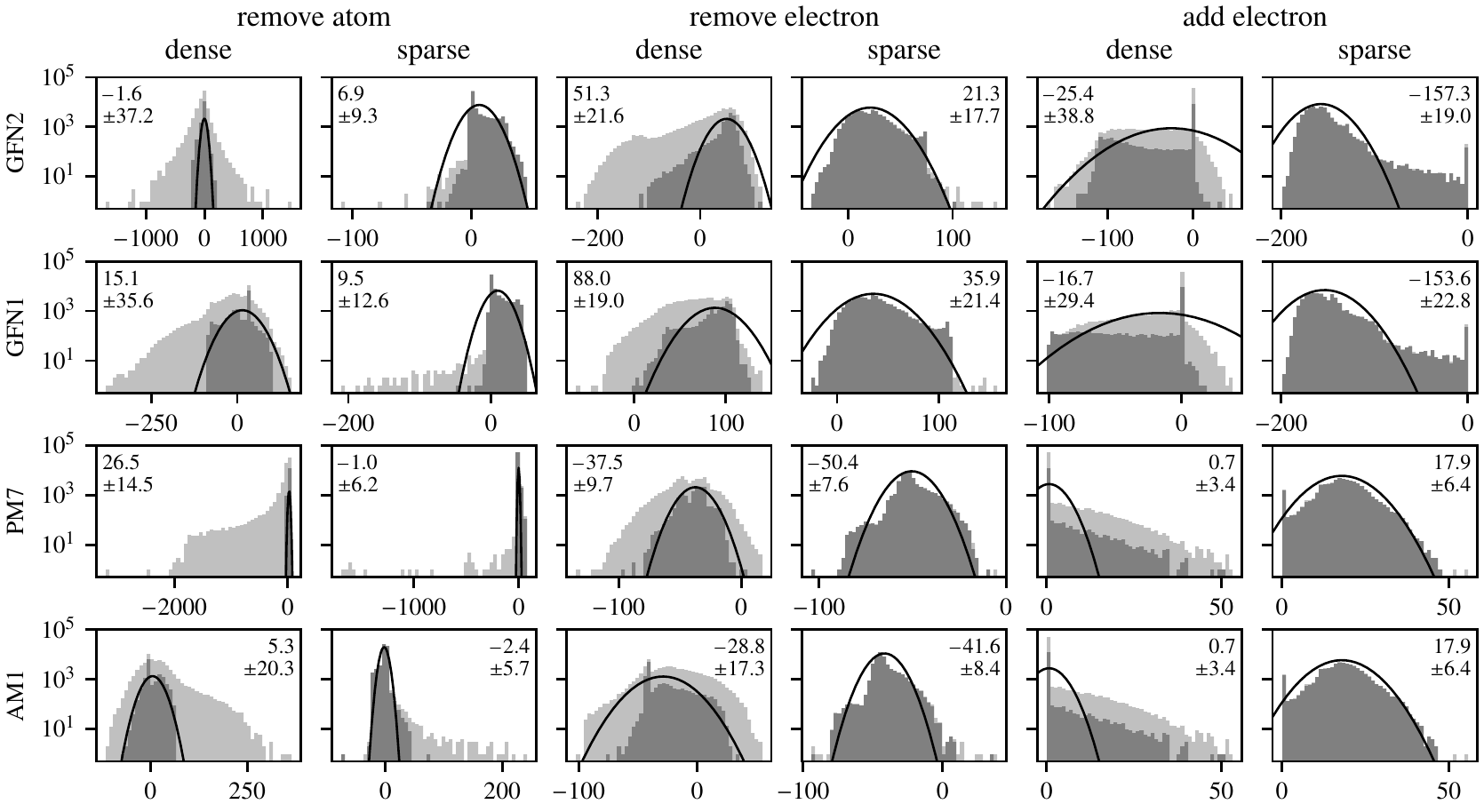}
    \caption{Error histograms in kcal/mol for SQM models and all tasks along with the means, standard deviations, and moment-matching Gaussian model fits of the marked data from structures with minimum interatomic distances greater than 0.74 {\AA}.}
    \label{distance_marked_plot}
\end{figure*}

Anomaly detection is a natural part of error analysis when gathering large amounts of data within a statistical framework.
The basic expectation of a good model is that its errors are an accumulation of a large number of small, independent errors, which tend to induce Gaussian distributions of model errors.
Errors in the hydrogen cluster data organized by model and task are shown in Fig.\@ \ref{error_plot} with moment-matching Gaussian fits.
While many errors are effectively described by the Gaussian model, there are also several fat error tails, many of which are rare enough to be unlikely to appear in data generation at smaller scales.
What is not shown are some even larger error tails that were present in earlier versions of the data set as the workflow was being refined to detect and avoid more failure events and silent errors.
This statistical overview of error distributions along with metadata collected during the primary data generation are essential for detecting and correcting rare failures.
Unfortunately, sufficiently rare failures are unlikely to occur in small-scale preliminary testing of a workflow precisely because of how rare they are.

There is not necessarily a clean partition between model, algorithm, and software errors in large-scale data generation.
For example, the lack of reliability in DIIS-based SCF solvers causes enough gaps in the ground-state searches that the wrong total spin is assigned in some DFT calculations.
As a result, some DFT calculations produce total energies that are too high, which are likely a source of some rare error outliers.
However, there is no guarantee that the DFT and HF ground states for a given structure and charge state will have the same total spin.
There is not enough information to distinguish model from algorithm errors here without more reliable SCF solver algorithms to fill gaps in data.
Similarly, software bugs may cause failures in one algorithm implementation that are not reproduced by other implementations, and custom improvements to algorithms may likewise cause successes that are also not reproducible in other software.
To show the impact of spin inconsistency, the DFT data is plotted in Fig.\@ \ref{spin_marked_plot} with spin-consistent calculations marked and fit to Gaussian error models.
The spin-inconsistent data contains most of the error outliers but does not substantially change the overall error statistics since the spin-consistent data has similar means and standard deviations.

The broadest error distributions in Fig.\@ \ref{error_plot} are in the SQM atom removal data from the dense distribution.
It is likely that errors in short-range pair potentials and matrix elements account for much of this error since these SQM models are mostly fit to data from near-equilibrium structures.
I test this hypothesis by separating data in Fig.\@ \ref{distance_marked_plot} based on the minimum interatomic distance in a structure being greater than or less than 0.74 {\AA}, the equilibrium bond length of H$_2$.
There is a clear narrowing of the error distributions for the structures without short interatomic distances, which supports the error hypothesis.

It may not be possible to detect or explain all error outliers.
The CCSD error tails from the sparse distribution in Fig.\@ \ref{error_plot} imply rare instances of large perturbative triples corrections to the total energy.
In these cases, the exact ground-state wave-function may have strong multi-reference character.
However, the multi-reference test in Fig.\@ \ref{rdm_validation} has no corresponding outliers, and a variety of multi-reference tests may be needed to increase detection reliability \cite{mr_tests}.

The failures that statistical model selection in Sec.\@ \ref{model_selection} seeks to avoid are silent failures.
Anomaly detection implies an ability to detect and herald some types of failures.
For the example data set in this paper, I chose to remove some heralded failures associated with algorithm-specific SCF convergence problems to increase the emphasis on errors in the physical models.
This formally changes the underlying task distributions by a small amount.
To be faithful to the original task distributions, a more complete model would always produce a viable output by branching to less accurate but more reliable calculations and eventually resorting to a random guess.
When trying to increase a model's overall success probability, improving the ability to detect and respond to rare failures and error outliers can be just as important as improving the average model accuracy for typical inputs.

\subsection{Model fitting}

I now consider a minimal representative example of using model selection to fit SQM models.
First, I highlight the benefits of using more complicated error models to improve success measures.
Second, I fit an atomic pair potential to all QM and SQM data, primarily to correct the large error outliers in the SQM data.
Pair potentials are one of the most common and basic elements of both interatomic potentials and SQM models.
While pair potentials are often restricted to a simple form before fitting them, I consider a general form and rely on model selection to limit the number of parameters that define the pair potential.

\begin{table}[!t]
  \centering
  \begin{tabular}{l c c l}
    \hline \hline
    model & $\hat{D}_{1\textrm{g}}$ & $\hat{D}_{6\textrm{g}}$ & $t$ (CPU-s) \\
    \hline
    CCSD+PP & $1.57 \times 10^5$ & $9.20 \times 10^4$ & $3.64 \times 10^8$ \\ 
    CCSD & $1.72 \times 10^5$ & $9.57 \times 10^4$ & $3.64 \times 10^8$ \\ 
    MP2 & $7.41 \times 10^5$ & $6.17 \times 10^5$ & $2.10 \times 10^8$ \\
    HF & $1.06 \times 10^6$ & $8.17 \times 10^5$ & $5.42 \times 10^7$ \\
    $\omega$B97M-V & $9.82 \times 10^5$ & $5.63 \times 10^5$ & $2.57 \times 10^8$ \\
    B3LYP & $1.09 \times 10^6$ & $6.28 \times 10^5$ & $8.51 \times 10^7$ \\
    PBE & $1.14 \times 10^6$ & $7.00 \times 10^5$ & $1.14 \times 10^8$ \\
    GFN2+PP & $1.53 \times 10^6$ & $1.17 \times 10^6$ & $9.31 \times 10^4$ \\
    GFN2 & $1.54 \times 10^6$ & $1.21 \times 10^6$ & $9.31 \times 10^4$ \\
    GFN1+PP & $1.52 \times 10^6$ & $1.13 \times 10^6$ & $9.08 \times 10^4$ \\
    GFN1 & $1.52 \times 10^6$ & $1.17 \times 10^6$ & $9.08 \times 10^4$ \\
    PM7+PP & $1.27 \times 10^6$ & $9.11 \times 10^5$ & $1.22 \times 10^6$ \\
    PM7 & $1.50 \times 10^6$ & $1.07 \times 10^6$ & $1.22 \times 10^6$ \\
    AM1+PP & $1.21 \times 10^6$ & $9.00 \times 10^5$ & $1.18 \times 10^6$ \\
    AM1 & $1.26 \times 10^6$ & $9.60 \times 10^5$ & $1.18 \times 10^6$ \\
    PP & $1.69 \times 10^6$ & $1.15 \times 10^6$ & $5.25 \times 10^{-2}$ \\
    none & $1.72 \times 10^6$ & $1.18 \times 10^6$ & $3.38 \times 10^{-2}$ \\
    \hline \hline
  \end{tabular}
  \caption{Comparison of minimized success measures over $m = 344,513$ simulation tasks for various models, including a pair potential (PP) correction when the improvement is greater than one percent.
  This comparison includes one-Gaussian (1g) and six-Gaussian (6g) error models.
  The total model evaluation times $t$ are in single-core CPU-seconds.
  The cost of generating the CCSD(T) reference data was $t = 4.13 \times 10^8$. \label{fit_table}
  }
\end{table}

The details of the model fitting process are as follows.
I used the success probability in Eq.\@ (\ref{exact_success}) for atom removal energies, Eq.\@ (\ref{signed_success}) for electron removal energies, and its non-positive version for electron addition energies.
For the goal of chemical accuracy, I used $\epsilon = 1$ kcal/mol.
I considered both a one-Gaussian error model with a common $\sigma$ and $\mu$ for all tasks and distributions, and a six-Gaussian error model with a different $\sigma$ and $\mu$ for each of the three tasks and two distributions.
The form of the pair potential was a polynomial at short range that goes to zero at an adjustable cutoff $R$ and strictly zero beyond that.
The success measure in Eq.\@ (\ref{success_measure}) and its analytical first and second derivatives with respect to $\boldsymbol{\lambda}$ were tedious but straightforward to evaluate.
I minimized the success measure with a sequence of line searches that used this derivative information to achieve asymptotic quadratic convergence.
As I increased the polynomial degree, I used the minimizing model with one fewer degree as the initial guess for minimization.
For a degree of one, I used moment-based approximations of $\sigma$ and $\mu$ from Fig.\@ \ref{error_plot} and a zero pair potential with $R = 4$ {\AA} as the initial guess.
Finally, the TIC bias correction in Eq.\@ (\ref{tic_bias}) was regularized by replacing small and negative eigenvalues of the Hessian with $10^{-9}$ times the largest eigenvalue when that value was greater.

The minimized success measures are summarized in Table \ref{fit_table}.
There is a clear benefit to using a richer error model with a separate Gaussian error model for each task and distribution.
As is clear from Fig.\@ \ref{error_plot}, the different tasks and distributions can have very different error models.
Forcing them all to have the same error model inevitably increases the overall standard deviation and reduces success probabilities.
Also, separate error models allow biases to be identified for individual tasks and distributions that would otherwise contribute to a larger standard deviation.
Some of these biases are obvious and expected, but it is still useful to quantify them.
The GFN1 and GFN2 models predict a very large electron binding energy for most hydrogen clusters, while AM1 and PM7 do not predict any binding of excess electrons to any hydrogen cluster.
The HF model has biases associated with the absence of electron correlation energy, which is always negative and usually proportional to the number of electrons.
The DFT models are known to have large delocalization errors \cite{delocal_dft} that are likely to be biasing the electron removal energies of the sparse distribution.
If these error models are used to improve success probabilities by adding random numbers to a model's outputs, then an improvement to the error model is an improvement to the model as a whole.

\begin{figure}[!b]
\includegraphics{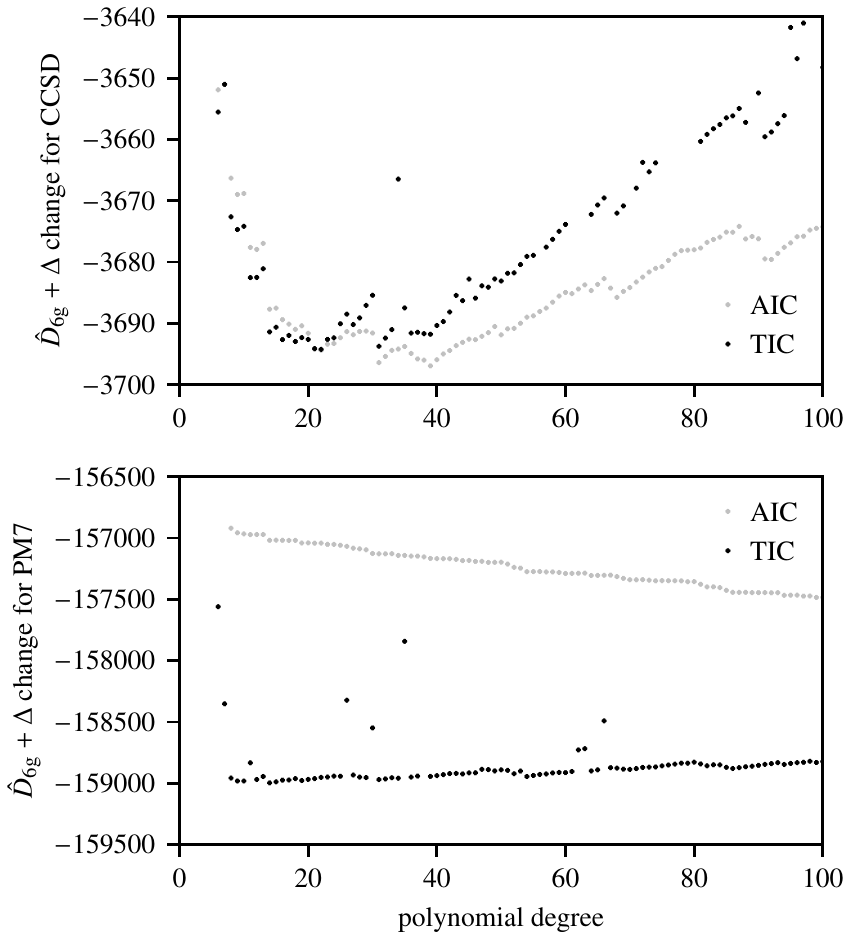}
\caption{Change in penalized success measures $\hat{D}_{\mathrm{6g}} + \Delta$ as the polynomial degree of the pair potential is increased, relative to having no pair potential.}
\label{poly_scan}
\end{figure}

The effects of the parameter penalties on the selection of the pair potential are shown for two representative model families in Fig.\@ \ref{poly_scan}.
CCSD is a more accurate model than PM7, and the AIC is likewise a better approximation of the TIC for CCSD.
For PM7, the AIC is unable to compensate for the parameter bias enough to create a local minimum in the success measure.
For CCSD, the AIC is able to create a local minimum, but its location is different than for the TIC.
In this example, the TIC correction introduces significant numerical noise even after regularization, which appear as values above the smoother trend line.
The TIC is a response property that depends sensitively on the numerical quality of the success measure minimum.
The derivative discontinuity that I allow at the large-distance cutoff point $R$ of the pair potential introduces derivative discontinuities in the $R$ dependence of the success measure that complicates the minimization.
Even under such non-ideal conditions, the TIC is still functional for model selection with appropriate regularization of the success measure Hessian.
The TIC is more challenging to calculate for parameterized QM calculations that involve QM response properties in the parameter derivatives of the success measure.

\begin{figure}[!t]
\includegraphics{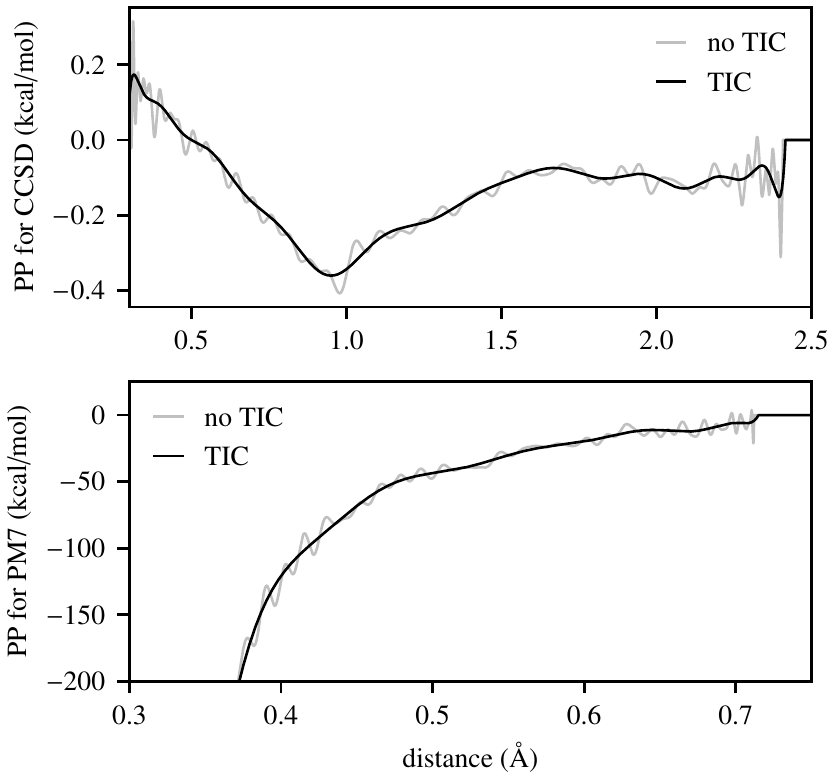}
\caption{Short-range polynomial pair potential corrections for the CCSD and PM7 models.
With the TIC penalty, the minimizing polynomial has degree 22 for CCSD and degree 14 for PM7.
Without the penalty, there is no local minimum in polynomial degree and best degree 100 polynomial is shown as an example of overfitting.
Outside of the plotted range, the PM7 pair potential decreases to -1790 kcal/mol at 0.3 {\AA}.}
\label{pair_pot}
\end{figure}

\begin{figure}[!b]
\includegraphics{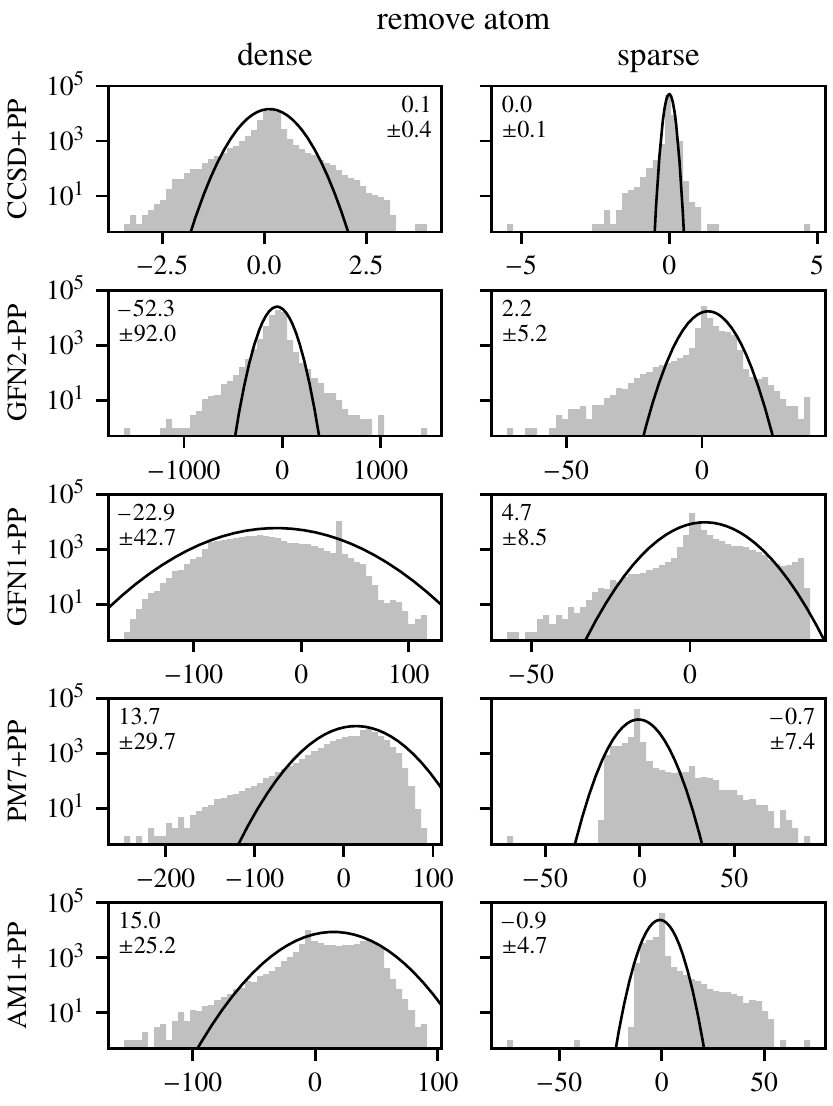}
\caption{Revisions of error histograms from Fig.\@ \ref{error_plot} in kcal/mol for the models and tasks that benefit from a pair potential correction.}
\label{fixed_error_plot}
\end{figure}

The benefits of a pair potential correction are not uniform over models or tasks.
Since the pair potential only depends on the atomic structure and not electronic structure, it cannot correct the electron addition and removal energies.
For many models, the overall reduction of the success measure is one percent or less, and these minor improvements are omitted from Table \ref{fit_table}.
The largest improvement comes from the PM7 pair potential, shown in Fig.\@ \ref{pair_pot}.
Apparently, the short-range hydrogen-hydrogen pair potential in PM7 is much too repulsive at distances just below the bond length of H$_2$.
In contrast, the CCSD pair potential is much longer in range and much smaller in magnitude.
It is not surprising that the largest correction occurs near the Coulson-Fischer point around 1 {\AA}.
However, it is surprising that something as complicated as the CCSD(T) triples correction can be partially approximated by a pair potential.
The parameter penalties succeed in suppressing the high-frequency oscillations typically attributed to overfitting noise, but there are still some artifacts near the edges of the polynomial's domain.
There are other ways to reduce unphysical oscillations in pair potentials, such as considering reference tasks that depend directly on derivatives of a pair potential or explicit functional regularization \cite{regularize_pot}.
As shown in Fig.\@ \ref{fixed_error_plot}, the pair potential corrections eliminate most of the large error outliers in SQM models except for GFN2 on the dense distribution.
I expect that the persistent error in GFN2 is from the short-range part of either a 3-body potential term or a Hamiltonian matrix element, neither of which can be repaired by a pair potential.
While many error tails have been eliminated by the pair potential correction, the corresponding improvements to the success measures in Table \ref{fit_table} are relatively modest.

This example does not demonstrate an efficient balance between the amount of data and the complexity of models.
There should be at least as many data points as model parameters to avoid trivial overfitting, and conventional wisdom in statistics suggests having at least an order of magnitude more data than parameters.
Practically, this corresponds to $\hat{D}(\boldsymbol{\lambda})$ being much larger than $\Delta$.
There is enough data here to support the fitting of models that are far more complicated than just a pair potential correction.
There may even be enough data to design and fit new low-complexity, many-parameter models that achieve chemical accuracy for hydrogen clusters.

\subsection{Cost budgeting\label{budgets}}

Considerations of model cost are always more complicated than model accuracy because they are much more sensitive to software, hardware, and fine details of a workflow.
All calculations reported in Table \ref{fit_table} were performed on the same computing cluster, with AMD EPYC 7702 CPU cores and two gigabytes of memory per core.
Except for MP2, CCSD, and CCSD(T) calculations, all calculations were performed on a single CPU core for maximum throughput.
Some of the MP2, CCSD, and CCSD(T) calculations exceeded the memory budget of a single CPU core, and they were run with four cores per calculation for a safety buffer of memory usage.
Parts of the calculation were threaded and made use of multiple cores, but the thread scaling was limited.
This complicates some cost comparisons.
For example, the HF and MP2 calculations had very similar run times under similar conditions, and the large cost difference reported in Table \ref{fit_table} is caused by the different number of cores required and poor thread scaling.
Also, the AM1 and PM7 calculations had similar run times as GFN1 and GFN2 calculations for individual calculations, but their workflow required a combinatorial search over atomic spin configurations.
The limited sensitivity of GFN1 and GFN2 calculations to spin order made this search unnecessary and reduced their overall run time per simulation task, but may also be related to their relatively poor accuracy here.

\begin{figure}[!t]
\includegraphics{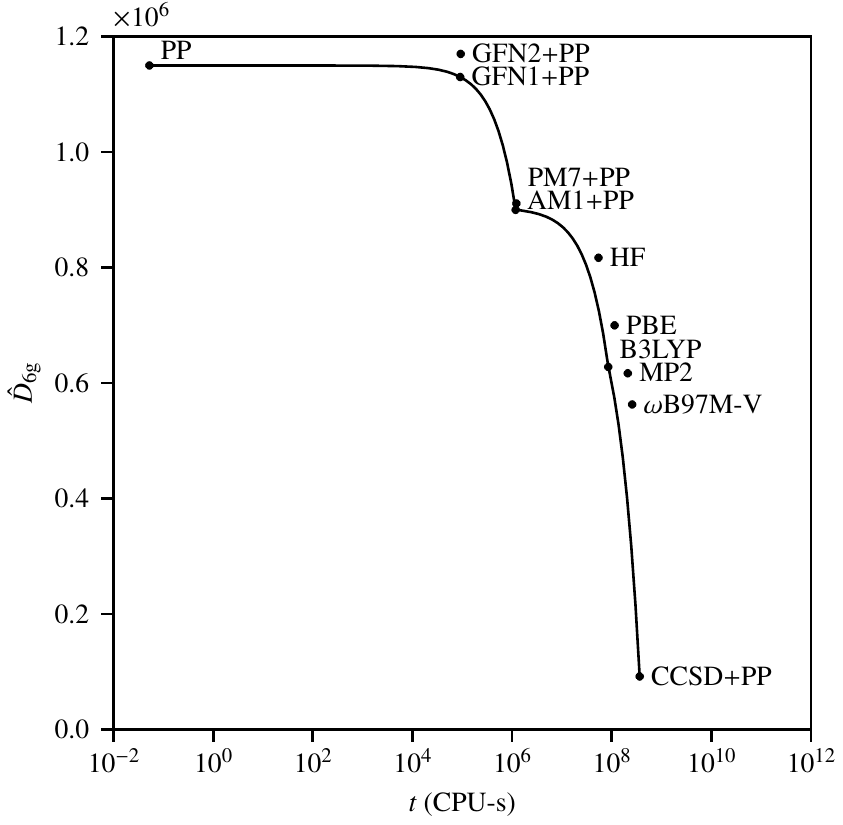}
\caption{Success measure versus total cost of models from Table \ref{fit_table} with the convex hull denoting the most cost-effective models for the example.}
\label{cost_plot}
\end{figure}

A visual way to compare success measures with varying cost penalties is to plot them versus cost as in Fig.\@ \ref{cost_plot}.
Following Eqs.\@ (\ref{mix_p}) and (\ref{mix_success}), models can be mixed randomly to vary cost and success continuously and satisfy expected budget constraints exactly.
This defines a convex hull of models with an optimal balance between cost and success, which is referred to as a Pareto front in multi-objective optimization \cite{moo}.
Models above the convex hull are not worth using for these simulation tasks according to this cost analysis.
In this example, the convex hull connects PP, AM1+PP, B3LYP, and CCSD+PP, with GFN1+PP also just on the convex hull.

Simultaneous considerations of model cost and accuracy at a large enough scale that reliability also matters as in Fig.\@ \ref{cost_plot} is a very challenging test for models.
It is much easier to show cost benchmarks of a model or software under ideal conditions, accuracy benchmarks under a different set of ideal conditions, and ignore problematic cases altogether.
Even the hydrogen cluster example considered here is artificially generous because a small fraction of structures that caused SCF convergence failures were omitted from the set of simulation tasks.
While the models are depicted as points on the plot, they are more generally going to be regions corresponding to the set of possible changes in a workflow that alter both cost and accuracy.
For example, the combinatorial search over atomic spin configurations for the hydrogen cluster example could have been avoided, which would have substantially reduced the cost of many models.
However, many of the calculations would have failed to find the lowest energy ground state, and the overall accuracy would have been reduced as a result.
Cost and accuracy could have been balanced more carefully by randomly sampling a limited set of spin configurations rather than using an exhaustive combinatorial search.
Adjusting details of a model workflow to improve the convex hull of optimal models requires a careful balance of these cost, accuracy, and reliability considerations.

\section{Discussion\label{discussion}}

There are three relevant topics of discussion adjacent to the results of this paper.
First, I discuss the variety of practical success probabilities that can be used within a statistical model selection framework.
Second, I discuss how a statistical framework relates to the concept of transferability in atomistic simulation.
Third, I discuss how the model quality observed in this paper might relate to practical and fundamental limits.

\subsection{Success probabilities\label{probability_choices}}

The use of a Gaussian error model for $z$ in Eq.\@ (\ref{success_target}) is justified by the central limit theorem if model errors come from an accumulation of many small, independent errors.
A non-zero mean further suggests that these small errors are biased on average.
However, since models do not have an infinite number of infinitesimal errors, the central limit theorem is only approximately valid and some errors in Eq.\@ (\ref{error_plot}) have non-Gaussian shapes.
It is possible to use other error models, and a better description can improve the success measure.
For example, a Laplace distribution of errors,
\begin{equation}
 \rho(z) = \frac{e^{-\sqrt{2}|z-\mu|/\sigma}}{\sigma \sqrt{2}},
\end{equation}
 would produce a success measure related to the MAE in the small-$\epsilon$ limit, similar to the relationship between a Gaussian error model and RMSD in Eq.\@ (\ref{rmsd_connection}).
More general probability distribution functions can be used, but they would likely require numerical integration to evaluate success probabilities.
MLE tends to match error models with observed errors.
However, the magnitude and correlation of errors is probably more important than the shape for improving model success.
By distinguishing when model errors are likely to be small or large \cite{atom_jackknife}, some high success probabilities can be maintained even when some success probabilities are inevitably low.
Ultimately, the choice of error model should be informed by the observed distribution of errors between model and data.

MLE is most effective when highly successful models exist and can be found.
However, this regime does not only apply to expensive models that achieve challenging success criteria such as chemically accurate energies.
It is also applicable to cheap models that are effective at easier tasks.
The concept of qualitative accuracy implies large but well-shaped errors that do not interfere with certain applications, and a more tolerant success criteria could be designed to match specific notions of qualitative accuracy.
For example, conformer searches only need to preserve the order of conformer energies, which can be tested by the Spearman rank correlation coefficient \cite{atom_statistics}.
Also, high-throughput computational searches often filter candidates through multiple layers of tests, with the cheapest tests applied to all candidates and more expensive tests applied only to candidates that pass the cheaper tests.
Models used for cheap tests should strongly prefer false positives that reduce filtering efficiency to false negatives that might accidentally eliminate good candidates.
An appropriate success criteria for these cheap models might then focus on the reduction of false negatives.

It is necessary for the success probabilities in Eq.\@ (\ref{success_probability}) to be statistically independent.
This can be problematic when models rely on some form of error correlation and cancellation between multiple calculations.
To accommodate this within a framework of independent success probabilities, the success criteria for a single simulation task must contain any comparisons between calculations that are needed.
In the simple example of a ranking task, random pairs of simulations might be performed with the output being the decision of which system had the larger value for a specific output.
With such a grouping, comparisons remain within a specific model even when different models are mixed as in Sec.\@ \ref{budgets} to control costs.
The practice of mixing models with different accuracies is usually avoided in atomistic simulation, but this framework provides a sensible approach to such mixing.

\subsection{Transferability}

A statistical framework for model selection can support more precise statistical statements about model transferability.
Here, I briefly contrast a notion of \textit{statistical transferability} from that of \textit{physical transferability}, which is frequently discussed when building models for atomistic simulation \cite{physical_transferability}.
I argue that while statistical transferability is the more desirable goal of model building, it is often impractical to avoid physical transferability assumptions given the present state of atomistic simulation methods.

Statistical transferability can directly predict the average future success of a model when simulation tasks can be interpreted as being drawn from the same distribution that was used to fit the model.
This is a form of model transferability to future simulation tasks that were not part of the reference data.
With enough reference data, $D(\boldsymbol{\lambda}^*)$ is approximately $\hat{D}(\boldsymbol{\lambda}^*) + \Delta$, and the average success probability per task from the distribution is approximately $\exp\{-[\hat{D}(\boldsymbol{\lambda}^*) + \Delta]/m\}$.
If the task distribution is designed to predict or approximate typical workloads of typical users of a model, then the model fitting process provides a direct operational statement about how effective the model should be for its users.

Statistical transferability can also be used to recycle reference data by transferring it between task distributions.
Reference data sampled from a second distribution $p'(X)$ over a superset of simulation tasks can be reused to approximate $\hat{D}(\boldsymbol{\lambda})$ for $p(X)$,
\begin{equation}
 \hat{D}(\boldsymbol{\lambda}) \approx - \left[ \min_i \frac{p'(X_i)}{p(X_i)} \right] \sum_{i=1}^m \frac{p(X_i)}{p'(X_i)} \log p(\boldsymbol{\lambda} | X_i).
\end{equation}
This is an implicit form of rejection sampling, and it requires the ability to calculate probability ratios between two task distributions.
It can also be used heuristically to reduce the influence of data that is necessary to fit a model but not representative of its typical applications.
Operationally, this can be interpreted as rare instances when users validate the model for themselves on the original reference data.
The effective sample size associated with this resampling procedure is
\begin{equation}
 m' = \left[ \min_i \frac{p'(X_i)}{p(X_i)} \right] \sum_{i=1}^m \frac{p(X_i)}{p'(X_i)},
\end{equation}
 which might be small if $p(X)$ and $p'(X)$ are very different.

In contrast, physical transferability is a set of observations and assumptions about the spatial locality of physics at an atomistic length scale.
It assumes that some model details and parameters describing short-range interatomic effects will be insensitive to distant changes in a large system with many atoms and then observes the varying degrees to which this is true.
The underlying first-principles QM equations are completely local and transferable when long-range interactions are mediated by local fields.
Unfortunately, locality and transferability are both degraded when encapsulating many-body effects and non-essential degrees of freedom to build simpler models.
Physical transferability assumptions are essential for justifying the use of methods that decompose large systems into a set of small fragments and simulate them individually, often embedded in simpler model environments.
Such methods include implicit solvation models \cite{implicit_solvent}, QM/MM embedding \cite{qmmm}, and the use of periodic supercells \cite{supercell}.
However, the effectiveness of these methods can be highly system dependent, an important example being the reduced locality of electronic effects in metallic systems that complicate efforts to develop low-cost methods \cite{linear_scaling}.

In the context of statistical model selection, physical transferability assumptions are unavoidable when generating reference data for task distributions containing large systems.
Reliable methods for reference data generation generally have large cost prefactors or poor cost scaling with system size that prevent their direct use on the task distribution.
Physical transferability can be used to justify the use of more accessible reference data corresponding to a proxy task distribution over small embedded fragments.
Tasks from the original distribution can be decomposed into sets of proxy tasks on fragments to generate the proxy distribution.
Similarly, cost models can be used to relate the observed simulation costs of fragments to the costs for larger systems.
While these small proxy tasks may all be contained within the original task distribution, the proxy distribution is over a strict subset of simulation tasks.
It is statistically impossible to sample from a distribution by weighting samples from a second distribution over a subset of events, but this is avoided by the physical fragmentation process.
While rigorous error analysis of this process is difficult, the general expectation is that the use of larger system fragments and more careful embedding increases the validity of physical transferability assumptions.

\subsection{Model quality}

In the limit of infinite effort spent on building atomistic models, Fig.\@ \ref{cost_plot} would represent a fundamental limit of cost versus accuracy for the example simulation tasks.
However, it is practically limited by finite effort and also reduced transferability of typical lower-cost models.
The hydrogen cluster example considered here is sufficiently different from typical reference data sets that it serves as a challenging test of physical transferability.
QM methods are less dependent on system-specific parameterizations and have naturally better physical transferability.
SQM methods might be able to improve their statistical transferability, but that would likely require substantially more data and parameterization than past model-building efforts.

Having an abundance of data creates a comfortable safety buffer between the number of parameters needed to fit a model accurately and the maximum number of parameters that can be fit with statistical significance.
The model selection process then enables an accurate model to be carved from an accessible set of redundant, overfit models.
Such large amounts of data are now accessible because of the massive scale of modern high-performance computing, an ability to generate data sets procedurally, and careful use of physical transferability assumptions.
This strongly contrasts with how SQM models such as PM7 \cite{pm7} and GFN2 \cite{gfn2} have been developed in the past.
They prescribed simple model forms with a few tens of parameters per element and collected enough reference data to fit those forms specifically.
They did not gather enough data to consider or rule out more complicated models with more parameters such as diatomic parameters, and many SQM model design choices have remained frozen for decades.
PM7 still uses the MNDO model form \cite{mndo} proposed in 1977, just with the addition of more complicated classical correction terms.
Despite being from a much newer model family, GFN2 also contains old model forms such as the Wolfsberg--Helmholz approximation \cite{wolfsberg_helmholz} from 1952 relating Hamiltonian and overlap off-diagonal matrix elements.
With an increasing amount of data, model forms can shift more towards what is objectively supported by the data and farther from the subjective technical opinions of specific model builders.

The total amount of development activity also impacts the progress of method development.
The development path of DFT from B3LYP to $\omega$B97M-V includes the development of hundreds of density functionals from numerous research groups over more than three decades \cite{mgcdb84}.
In contrast, the development path from AM1 to PM7 consists of only a few other models developed by a single scientist -- Dr. James J. P. Stewart -- working mostly in isolation outside of academia for more than three decades.
GFN1 and GFN2 were developed much more recently by a single academic group -- the research group of Prof. Stefan Grimme at the University of Bonn.
While there are other SQM models outside of the GFN and MNDO-like model families, these are the two most widely used families and the only non-commerical models \cite{quasinano} to be fit for combinations of elements over most of the periodic table.
The GFN models incorporate ideas from both MNDO-like models (multipole expansions of electrostatics, avoidance of diatomic parameters) and DFTB models (expansion around an atomic limit, DFT-like correlation models).
All of the SQM models considered here have similar superficial complexity, similar numbers of parameters per element, and are fit to similar amounts of reference data.
While likely a coincidence, these SQM models have systematically degrading performance on the hydrogen cluster example in chronological order of their development.
The present limitations of SQM models may reflect a limited investment in their development more than any fundamental limitation.

Machine learning (ML) models \cite{ml_review} are another popular class of atomistic models that tend to be cheaper and more heavily parameterized than SQM models.
ML models can also be combined with SQM models \cite{ml_sqm}.
No ML models were considered in this paper because there were no ML models available with parameters appropriate for the hydrogen cluster example.
Statistical model selection is certainly applicable to ML models, and parameter penalties are particularly relevant since ML models often have a large number of parameters.
ML model families are often designed to be universal approximants in the limit of an infinite number of parameters, which could allow them to span the whole cost versus accuracy range of Fig.\@ \ref{cost_plot}.
Where ML models contribute to the Pareto front for a particular task distribution is likely to depend on how much they are developed relative to QM and SQM models and how efficiently they utilize their parameters.
However, the common practice of overfitting ML models \cite{ml_overfit} is not supported by statistical model selection and will cause the parameter penalty in Eq.\@ (\ref{tic_bias}) to diverge when $\mathbf{F}$ develops a robust null space.

\section{Conclusion}

The concepts presented in this paper are meant to inform the process of designing, fitting, and selecting models for atomistic simulation tasks.
If simulation tasks are not going to be repeated a very large number of times, then the formal process of gathering reference data and calculating a success measure might not be worth the amount of human effort required.
However, the statistical model selection process can still be useful as a conceptual guide even when it is not worthwhile to perform it carefully or explicitly.
For tasks that are performed frequently by many scientists, it may be worthwhile to capture that activity as a distribution of tasks and a representative sampling from that distribution.
Quantum chemistry has a tradition of curating reference data sets to guide method development \cite{mgcdb84}.
Expanding that tradition to accommodate larger data sets, statistical interpretations, and success measures that capture the real needs of applied scientists could create an even better guide for method development.
It is difficult for a scientist to characterize real application needs while also developing novel simulation methods to satisfy those needs, and it would be helpful to decouple those important research activities from each other.

An essential aspect of model building in atomistic simulation is the availability of high-quality reference data for fitting and testing.
While models have historically relied on reference data from experiments, it is now possible to generate accurate data using expensive QM models.
As shown in the hydrogen cluster example, CCSD(T) data is affordable for small molecular fragments, and less accurate DFT data remains affordable for larger molecules and periodic systems.
For data generation at scales larger than what has been presented in this paper, reliability issues will become increasingly important alongside cost and accuracy considerations.
SCF convergence problems can cause heralded failures, while SCF convergence to excited states can cause silent failures.
Without more fundamentally reliable algorithms to reduce failure rates, a fixed rate of failure means an increasing number of failure events as data sets grow larger in size.
There are increasingly sophisticated tools \cite{qcarchive} for remote, automated computing of large workloads and organizing large data sets with modern database standards.
However, limitations in the reliability of the underlying tasks being automated may have a strongly negative influence on the cost and accuracy of generating large data sets as failures persist against increasing computational redundancy.

As scientists continue to develop more diverse and sophisticated models for atomistic simulation, how models are compared and how their successes are judged will become increasingly important.
Progress in method development can slow down or stop if scientists have different, incompatible definitions for what success is \cite{dft_density}.
This paper has presented an operational success measure for judging atomistic models that is based on statistical model selection.
Using simple simulation tasks on hydrogen clusters as an example, I have shown how this measure can be used to compare the cost and accuracy of a diverse set of QM and SQM models.
I have also used it to fit minimal SQM models that apply pair potential corrections to this QM and SQM data and selected the potential form that best fit the data.
The TIC provides a reliable parameter penalty to avoid selecting over-complicated models, while the AIC is not a reliable penalty because some atomistic models are too inaccurate for its assumptions to hold.
For a computational budget that is too small for a high-accuracy model but excessive for a low-accuracy model, the success measure predicts the efficacy of splitting a workload between models to match the budget.
By adjusting the operational definition of success for simulation tasks, this success measure can be equally good for designing expensive models to succeed at difficult tasks and cheap models to succeed at easy tasks.
In either case, this paper prescribes a process for fitting and choosing the model with the highest chance of success.


\begin{acknowledgments}
J.\@ E.\@ M.\@ thanks Jimmy Stewart for helpful discussions.
The Molecular Sciences Software Institute is supported by NSF Grant No.\@ ACI-1547580.
The computational resources used in this work were provided by Advanced Research Computing at Virginia Tech.
\end{acknowledgments}

\section*{Author declarations}

\subsection*{Conflict of Interest}

The author has no conflicts to disclose.

\subsection*{Author Contributions}

\noindent \textbf{Jonathan E. Moussa:} Conceptualization (equal); Data curation (equal); Formal analysis (equal); Investigation (equal); Methodology (equal); Resources (equal); Software (equal); Validation (equal); Visualization (equal); Writing -- original draft (equal); Writing -- review \& editing (equal).

\section*{Data Availability}

The data and software that support the findings of this study are available on Zenodo at the DOI \href{https://doi.org/10.5281/zenodo.7530231}{10.5281/zenodo.7530231}.

\appendix

\section{DERIVATION OF INFORMATION CRITERIA \label{ic_derivation}}

While practical model selection works with a specific reference data set, the derivation of Eqs.\@ (\ref{aic_bias}) and (\ref{tic_bias}) requires averaging over reference data sets.
To clarify this with the notation, I label $\hat{D}(\boldsymbol{\lambda})$ here with a specific data set $\mathbf{X}$ as
\begin{equation}
 \hat{D}_{\mathbf{X}}(\boldsymbol{\lambda}) = - \sum_{i=1}^m \log p(\boldsymbol{\lambda} | X_i).
\end{equation}
I also use a compact notation for averages over reference data sets,
\begin{equation}
 \overline{\sum_{\mathbf{X}}} = \sum_{X_1} p(X_1) \cdots \sum_{X_m} p(X_m),
\end{equation}
which simplifies many of the equations presented here.
However, all $\mathbf{X}$ subscripts are removed when referring to these results in the main text, causing a small inconsistency in notation.

The premise of both the AIC and TIC is that $D(\boldsymbol{\lambda})$ is free of finite-sampling errors that distort the parameter optimization process.
The goal of their parameter penalties is to approximate the minimum value of $D(\boldsymbol{\lambda})$ with properties of $\hat{D}(\boldsymbol{\lambda})$, since $D(\boldsymbol{\lambda})$ is not efficient to compute directly.
I refer here to the minimizer of $D(\boldsymbol{\lambda})$ as $\boldsymbol{\lambda}^*$ and the minimizers of $\hat{D}_{\mathbf{X}}(\boldsymbol{\lambda})$ as $\boldsymbol{\lambda}^*_{\mathbf{X}}$.

For a constant value of $\boldsymbol{\lambda}$, $\hat{D}_{\mathbf{X}}(\boldsymbol{\lambda})$ is an unbiased estimator of $D(\boldsymbol{\lambda})$ when averaged over sets of $m$ simulation tasks $\mathbf{X}$,
\begin{equation} \label{average_relation}
  D(\boldsymbol{\lambda}) = \overline{\sum_{\mathbf{X}}} \hat{D}_{\mathbf{X}}(\boldsymbol{\lambda}).
\end{equation}
Since I cannot efficiently calculate $\boldsymbol{\lambda}^*$, I would like to evaluate $D(\boldsymbol{\lambda})$ at one $\boldsymbol{\lambda}^*_{\mathbf{X}}$ value that I can calculate efficiently.
If this was repeated and averaged over sets of $m$ simulation tasks, it would be an unbiased estimator of
\begin{equation} \label{ic_target}
  \overline{D}_{\mathrm{min-ave}} = \overline{\sum_{\mathbf{X}}} D(\boldsymbol{\lambda}^*_{\mathbf{X}}).
\end{equation}
However, with a single $\mathbf{X}$, I can only evaluate $\hat{D}_{\mathbf{X}}(\boldsymbol{\lambda})$ at its own minimum $\boldsymbol{\lambda}^*_{\mathbf{X}}$, which is an unbiased estimator of the average minimum,
\begin{equation}
  \overline{D}_{\mathrm{ave-min}} = \overline{\sum_{\mathbf{X}}} \hat{D}_{\mathbf{X}}(\boldsymbol{\lambda}^*_{\mathbf{X}}).
\end{equation}
This has a negative bias relative to Eq.\@ (\ref{ic_target}) because each $\hat{D}_{\mathbf{X}}(\boldsymbol{\lambda})$ is evaluated at its own minimum instead of a common $\boldsymbol{\lambda}$.
A single $\hat{D}_{\mathbf{X}}(\boldsymbol{\lambda}^*_{\mathbf{X}})$ value can be unbiased as an estimator of $D(\boldsymbol{\lambda}^*_{\mathbf{X}})$ by adding a bias correction,
\begin{equation}
 \Delta = \overline{D}_{\mathrm{min-ave}} - \overline{D}_{\mathrm{ave-min}} = \overline{\sum_{\mathbf{X}}} [ D(\boldsymbol{\lambda}^*_{\mathbf{X}}) - \hat{D}_{\mathbf{X}}(\boldsymbol{\lambda}^*_{\mathbf{X}}) ].
\label{define_bias}
\end{equation}
I approximate $\Delta$ after several simplifying assumptions.

The first assumption is that $D(\boldsymbol{\lambda})$ and $\hat{D}_{\mathbf{X}}(\boldsymbol{\lambda})$ are both slowly changing in a region containing $\boldsymbol{\lambda}^*$ and $\boldsymbol{\lambda}^*_{\mathbf{X}}$.
Both functions can be extrapolated from their minimum to the other function's minimum with a second-order Taylor expansion,
\begin{align}
 D(\boldsymbol{\lambda}^*_{\mathbf{X}}) &\approx D(\boldsymbol{\lambda}^*) + \frac{1}{2} (\boldsymbol{\lambda}^*_{\mathbf{X}} - \boldsymbol{\lambda}^*)^T \mathbf{F} (\boldsymbol{\lambda}^*_{\mathbf{X}} - \boldsymbol{\lambda}^*), \notag \\
 \hat{D}_{\mathbf{X}}(\boldsymbol{\lambda}^*) &\approx \hat{D}_{\mathbf{X}}(\boldsymbol{\lambda}^*_{\mathbf{X}}) + \frac{1}{2} (\boldsymbol{\lambda}^* - \boldsymbol{\lambda}^*_{\mathbf{X}})^T \mathbf{F}_{\mathbf{X}} (\boldsymbol{\lambda}^* - \boldsymbol{\lambda}^*_{\mathbf{X}}), \notag \\
   [\mathbf{F}]_{i,j} &= \frac{\partial^2 D}{\partial\lambda_i \partial\lambda_j}(\boldsymbol{\lambda}^*), \ \ \ [\mathbf{F}_{\mathbf{X}}]_{i,j} = \frac{\partial^2 \hat{D}_{\mathbf{X}}}{\partial\lambda_i \partial\lambda_j}(\boldsymbol{\lambda}^*_{\mathbf{X}}).
\end{align}
These extrapolations can be combined using Eq.\@ (\ref{average_relation}) to simplify the bias correction in Eq.\@ (\ref{define_bias}) to
\begin{equation}
 \Delta \approx  \frac{1}{2} \overline{\sum_{\mathbf{X}}} (\boldsymbol{\lambda}^* - \boldsymbol{\lambda}^*_{\mathbf{X}})^T (\mathbf{F} + \mathbf{F}_{\mathbf{X}}) (\boldsymbol{\lambda}^* - \boldsymbol{\lambda}^*_{\mathbf{X}}).
\label{simpler_bias}
\end{equation}
Similarly, I can extrapolate $\hat{D}_{\mathbf{X}}(\boldsymbol{\lambda})$ from $\boldsymbol{\lambda}^*$ to $\boldsymbol{\lambda}^*_{\mathbf{X}}$,
\begin{align}
   \hat{D}_{\mathbf{X}}(\boldsymbol{\lambda}^*_{\mathbf{X}}) &\approx \hat{D}_{\mathbf{X}}(\boldsymbol{\lambda}^*) + (\boldsymbol{\lambda}^*_{\mathbf{X}} - \boldsymbol{\lambda}^*)^T \frac{\partial \hat{D}_{\mathbf{X}}}{\partial\boldsymbol{\lambda}}(\boldsymbol{\lambda}^*) \notag \\
  & \ \ \ + \frac{1}{2} (\boldsymbol{\lambda}^*_{\mathbf{X}} - \boldsymbol{\lambda}^*)^T \mathbf{F}'_{\mathbf{X}} (\boldsymbol{\lambda}^*_{\mathbf{X}} - \boldsymbol{\lambda}^*),
 \notag \\
  [\mathbf{F}'_{\mathbf{X}}]_{i,j} &= \frac{\partial^2 \hat{D}_{\mathbf{X}}}{\partial\lambda_i \partial\lambda_j}(\boldsymbol{\lambda}^*),
\end{align}
 and minimize the quadratic form for the parameter variations,
\begin{equation}
 \boldsymbol{\lambda}^* - \boldsymbol{\lambda}^*_{\mathbf{X}} \approx (\mathbf{F}'_{\mathbf{X}})^{-1} \frac{\partial \hat{D}_{\mathbf{X}}}{\partial\boldsymbol{\lambda}}(\boldsymbol{\lambda}^*) .
\label{variation_solve}
\end{equation}
The second assumption is that $\mathbf{F} \approx \mathbf{F}_{\mathbf{X}} \approx \mathbf{F}'_{\mathbf{X}}$, which allows for the removal of $\mathbf{F}_{\mathbf{X}}$ and $\mathbf{F}'_{\mathbf{X}}$ from Eq.\@ (\ref{simpler_bias})
 after substituting Eq.\@ (\ref{variation_solve}),
\begin{align}
  \Delta &\approx \mathrm{tr} [ \mathbf{F}^{-1} \mathbf{\tilde{F}} ], \notag \\
  [\mathbf{\tilde{F}}]_{i,j} &= \overline{\sum_{\mathbf{X}}} \frac{\partial \hat{D}_{\mathbf{X}}}{\partial \lambda_i}(\boldsymbol{\lambda}^*) \frac{\partial \hat{D}_{\mathbf{X}}}{\partial\lambda_j}(\boldsymbol{\lambda}^*) .
\label{simple_bias}
\end{align}
The validity of these two assumptions can be increased by adding more reference data to reduce finite-sample effects until
 $\hat{D}_{\mathbf{X}}(\boldsymbol{\lambda})$ and $D(\boldsymbol{\lambda})$ have small differences in their gradients and negligible differences in their Hessians at $\boldsymbol{\lambda}^*_{\mathbf{X}}$ and $\boldsymbol{\lambda}^*$.

The TIC follows from a related assumption about small finite-sampling effects.
As a useful reference, I rearrange $\mathbf{F}$ and $\mathbf{\tilde{F}}$ into a similar form by rewriting $\mathbf{\tilde{F}}$ as a sum over simulation tasks rather than over groups of $m$ simulation tasks,
\begin{align}
 [\mathbf{\tilde{F}}]_{i,j} &= m \sum_{X} p(X) \left[ \frac{\partial \log p(\boldsymbol{\lambda} | X)}{\partial\lambda_i} \frac{\partial \log p(\boldsymbol{\lambda} | X)}{\partial\lambda_j} \right]_{\boldsymbol{\lambda} = \boldsymbol{\lambda}^*}, \notag \\
 [\mathbf{F}]_{i,j} &= - m \sum_{X} p(X) \left[ \frac{\partial^2 \log p(\boldsymbol{\lambda} | X)}{\partial\lambda_i \partial\lambda_j} \right]_{\boldsymbol{\lambda} = \boldsymbol{\lambda}^*}.
 \label{fisher_info}
\end{align}
The TIC bias correction is a direct approximation of Eq.\@ (\ref{simple_bias}) by
\begin{align}
  \Delta &\approx \Delta_{\mathrm{TIC}} = \mathrm{tr} [ \mathbf{F}_{\mathbf{X}}^{-1} \mathbf{\tilde{F}}_{\mathbf{X}} ], \notag \\
  [\mathbf{\tilde{F}}_{\mathbf{X}}]_{i,j} &= \sum_{k=1}^m \left[ \frac{\partial \log p(\boldsymbol{\lambda} | X_k)}{\partial\lambda_i} \frac{\partial \log p(\boldsymbol{\lambda} | X_k)}{\partial\lambda_j} \right]_{\boldsymbol{\lambda} = \boldsymbol{\lambda}^*_{\mathbf{X}}},
\label{tic_bias2}
\end{align}
 which again assumes that the $m$ samples in $\mathbf{X}$ are sufficient to converge expectation values so that $\mathbf{\tilde{F}} \approx \mathbf{\tilde{F}}_{\mathbf{X}}$ and $\mathbf{F} \approx \mathbf{F}_{\mathbf{X}}$.

The AIC follows from additional assumptions about model accuracy.
I can simplify the difference between $\mathbf{F}$ and $\mathbf{\tilde{F}}$ in Eq.\@ (\ref{fisher_info}) by rearranging and combining the logarithmic derivatives into
\begin{equation} \label{fisher_residual}
  [\mathbf{\tilde{F}} - \mathbf{F}]_{i,j} = m \sum_{X} \frac{p(X)}{p(\boldsymbol{\lambda}^* | X)} \left[ \frac{\partial^2 p(\boldsymbol{\lambda} | X)}{\partial\lambda_i \partial\lambda_j} \right]_{\boldsymbol{\lambda} = \boldsymbol{\lambda}^*}.
\end{equation}
Next, I consider a modified form of $D(\boldsymbol{\lambda})$ from Eq.\@ (\ref{divergence}) in which the reference simulation tasks are assigned a failure rate $\delta$,
\begin{align}
   D(\boldsymbol{\lambda}) &= - m \sum_{X} (1-\delta) p(X) \log p(\boldsymbol{\lambda} | X) \notag \\
    & \ \ \ - m \sum_{X} \delta p(X) \log (1 - p(\boldsymbol{\lambda} | X)).
\end{align}
The original form is recovered in the $\delta \rightarrow 0$ limit.
If the derivation is repeated for the modified form, Eq.\@ (\ref{fisher_residual}) becomes
\begin{align}
  [\mathbf{\tilde{F}} - \mathbf{F}]_{i,j} &= m \sum_{X} \frac{(1-\delta) p(X)}{p(\boldsymbol{\lambda}^* | X)} \left[ \frac{\partial^2 p(\boldsymbol{\lambda} | X)}{\partial\lambda_i \partial\lambda_j} \right]_{\boldsymbol{\lambda} = \boldsymbol{\lambda}^*} \notag \\
  & + m \sum_{X} \frac{\delta p(X)}{1-p(\boldsymbol{\lambda}^* | X)} \left[ \frac{\partial^2 [1 - p(\boldsymbol{\lambda} | X)]}{\partial\lambda_i \partial\lambda_j} \right]_{\boldsymbol{\lambda} = \boldsymbol{\lambda}^*}.
\label{fisher_residual2}
\end{align}
The final AIC assumption is that the optimized model can recover the reference distribution, resulting in $p(\boldsymbol{\lambda}^* | X) \approx 1 - \delta$ here.
I can then cancel the $\delta$ factors and combine the two terms in Eq.\@ (\ref{fisher_residual2}),
\begin{equation}
  [\mathbf{\tilde{F}} - \mathbf{F}]_{i,j} \approx m \left[ \frac{\partial^2}{\partial\lambda_i \partial\lambda_j} \sum_{X} p(X)\right]_{\boldsymbol{\lambda} = \boldsymbol{\lambda}^*} = 0.
\end{equation}
The difference between $\mathbf{\tilde{F}}$ and $\mathbf{F}$ disappears for any value of $\delta$.
The AIC bias correction corresponds to ignoring this difference and keeping only the trace of the identity matrix over the $n$-dimensional parameter space,
\begin{equation} \label{aic_correction}
 \Delta = n + \mathrm{tr} [ \mathbf{F}^{-1} (\mathbf{\tilde{F}} - \mathbf{F}) ] \approx \Delta_{\mathrm{AIC}} = n.
\end{equation}
The validity of the good model assumption can be increased by improving the model family and relaxing the success criteria to increase all optimized success probabilities towards one.

\section{DETAILS OF DATA GENERATION\label{data_generation}}

All QM calculations used a post-2.1.1 development version of PySCF \cite{pyscf1,pyscf2,pyscf_mod}.
All calculations used spin-unrestricted orbitals.
For HF theory and every DFT functional, the large-basis calculations were initialized by projecting a converged density matrix from a calculation in the smaller def2-SVP basis set \cite{qzvpp}.
The def2-SVP density matrix was taken from the calculation with the lowest total energy from a systematic ground-state search for each structure and charge state.
First, a def2-SVP calculation was performed for every spin state from the standard spin-averaged independent-atom density matrix guess.
Second, a custom density matrix guess was constructed from spin-polarized independent-atom density matrices with every combination of atomic charges and spin orientations.
Third, after performing all of these small-basis calculations with the default DIIS algorithm \cite{diis}, they were all repeated with an alternative ADIIS algorithm \cite{adiis}.
The large-basis calculation used the same algorithm, either DIIS or ADIIS, as the small-basis calculation that was used to initialize it.
Even with all of this redundancy, it was not possible to converge the SCF cycle for every charge and spin state of every structure.
While the variational nature of SCF calculations guarantees the existence of stable local energy minima, DIIS-based algorithms provide no guarantees of convergence to them.
All large-basis DFT calculations use a (99,590) local grid and a SG-1 nonlocal grid, following the recommendations for the $\omega$B97M-V functional \cite{wb97mv}.

For SQM calculations, MOPAC 22.0.5 \cite{mopac} was used for AM1 and PM7 calculations, and xTB 6.5.1 \cite{xtb} was used for GFN1 and GFN2 calculations.
MOPAC calculations followed the same ground-state search procedure as the PySCF calculations except with only DIIS and without any projection into a larger basis.
There are fewer points of failure in minimal-basis calculations, and MOPAC was able to converge an SCF calculation for every structure and charge state.
xTB calculations do not contain Fock exchange and depend on an initial electronic density guess rather than a density matrix guess.
I only used the default spin-averaged density guess and restricted the ground-state search to total spin values.
There was a high failure rate for SCF convergence in xTB with the default options for this data set.
However, it was possible to converge every structure and charge state in xTB with calculations at elevated electronic temperatures of 3000 K and then 1500 K followed by linear extrapolation of the total energies to zero temperature.

At this level of automation and scale of data generation, it was not possible to converge every iterative solve for HF, DFT, and CCSD calculations in PySCF.
The choice of solver options was important as it changed success statistics and average run times.
I did not try to optimize these choices in a systematic way, but they were adjusted during the implementation of the workflow to improve success rates \cite{fine_details}.
In addition to convergence failures, a DFT or HF calculation was considered to fail if the def2-QZVPP total energy was more than 10 kcal/mol larger than the smallest def2-SVP total energy.
Total energies tend to be lower for larger basis sets because they have more variational degrees of freedom.
I attribute these energy increases to the DIIS phenomenon of escaping from the basin of convergence of a ground state and converging to a very different stationary state with a larger energy.
The failure rate of DFT calculations was 3.4\%, the failure rate of CCSD(T) calculations was 1.0\%, and the overall failure rate of the simulation tasks was 4.3\%.
If at least one model failed to produce an output for a simulation task, then that task was omitted from the final data set and statistical analysis.
Such failures distort the distribution of simulation tasks because they act as a form of rejection sampling.

\begin{figure}
\includegraphics{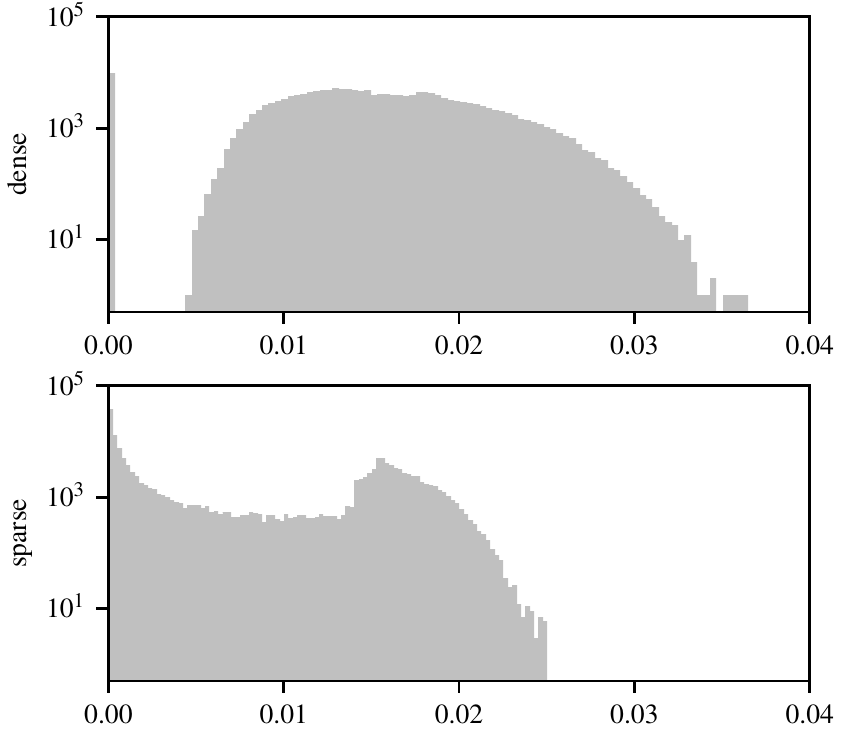}
\caption{Histograms of the maximum deviation from zero and one of the unrelaxed MP2 1RDM eigenvalues for all structures and charge states from the reference data sets.
Strong multi-reference effects would correspond to values near 0.5, which are not observed in this data set.}
\label{rdm_validation}
\end{figure}

I also validated the CCSD(T)/def2-QZVPP level of theory for this data set while gathering data.
The main validity concern is strong electron correlation effects, which are known to occur in hydrogen clusters \cite{hydrogen_chain}.
These effects are caused by multi-reference ground states that come from a superposition of many electronic spin configurations with nearly degenerate energies in the atomic limit.
Randomly generated hydrogen clusters are unlikely to have many degenerate spin configurations, and they are expected to be more weakly correlated on average.
The most direct validity test would be the overlap between the normalized HF and CCSD many-body wave-functions, but this quantity is not efficiently computable.
Instead, I used the eigenvalues of the one-particle density matrix (1RDM) at the unrelaxed MP2 level of theory as an accessible proxy for this overlap.
The maximum deviation of the eigenvalues from zero and one is strictly zero when the overlap is one, and the deviation increases as the overlap is reduced.
This deviation is plotted for every structure in every charge state in Fig.\@ \ref{rdm_validation}.
The sparse distribution that is expected to be more susceptible to multi-reference effects because of spin symmetry breaking does not have larger deviations than the dense distribution.

The other major validity concern is the basis-set convergence of CCSD(T)/def2-QZVPP.
A quadruple-zeta basis such as def2-QZVPP does not typically converge absolute post-HF energies to chemical accuracy of 1 kcal/mol or less without basis-set extrapolation or explicit correlation corrections.
However, the simulation tasks considered here only require energy differences between structures that differ by at most one atom, which should be less sensitive to finite-basis errors.
The most basis-set sensitive structures are correlation-bound anions, which account for 5.3\% of the anions in the dense distribution and 45.2\% in the sparse distribution.
Correlation-bound anions do not have a proper complete basis set (CBS) limit with HF orbitals because the overlap between the HF and CCSD wave-functions tends to zero as the unbound HF orbital delocalizes further into the vacuum.
A formally correct treatment of correlation-bound anions in the CBS limit requires Brueckner orbitals \cite{cba}.
However, I do not expect the def2-QZVPP basis set to be large enough for the pathological CBS limit to have a substantial effect on this data set.
The success in projecting HF and DFT solutions from def2-SVP to def2-QZVPP was empirical evidence in support of this expectation.


\begin{thebibliography}{99}
\bibitem{pyscf_mod} Several quantum chemistry codes were considered for this work, but I was not able to get all of the necessary methods for this project working in any one code without modifications.
I ultimately chose PySCF because the open-source Python codebase made it easier to find and fix bugs and contribute bug fixes.
The offending bugs were associated with logical problems when a fully spin-polarized system had no beta electrons in some post-HF methods and a memory leak in the CCSD(T) code that caused crashes in long workflows.
\bibitem{fine_details} The most relevant details for success rates were SCF convergence tolerances set to $10^{-8}$, the maximum number of DIIS vectors set to $10$, and the maximum number of SCF cycles set to 100 for def2-SVP and 200 for def2-QZVPP.
The CCSD calculations were also set to a maximum number of 200 iterations and a convergence tolerance of $10^{-5}$ on the cluster operator.
\end{thebibliography}
\end{document}